\newcommand{\diracslash}[1]{#1\llap{/\kern2pt}}

\newcommand{\be}{\begin{equation}}
\newcommand{\ee}{\end{equation}}
\newcommand{\bea}{\begin{eqnarray}}
\newcommand{\eea}{\end{eqnarray}}
\newcommand{\ba}[1]{\begin{array}{#1}}
\newcommand{\ea}{\end{array}}

\newcommand{\bt}{\begin{tabular}}
\newcommand{\et}{\end{tabular}}

\newcommand{\beas}{\begin{eqnarray*}}
\newcommand{\eeas}{\end{eqnarray*}}

\documentclass[preprint,prd,aps,floats,nofootinbib,floatfix]{revtex4}

\DeclareSymbolFont{rsfs}{U}{rsfs}{m}{n}
\DeclareSymbolFontAlphabet{\mathrsfs}{rsfs}

\usepackage{graphicx}

\usepackage{multirow}
\usepackage{graphicx,epstopdf}

\usepackage{graphicx}
\usepackage{subcaption}
\usepackage[utf8]{inputenc}
\usepackage{csquotes}
\usepackage[english]{babel}
\usepackage{hyperref}
\hypersetup{
    colorlinks=true,
    linkcolor=blue,
    filecolor=magenta,     
    citecolor=blue, 
    urlcolor=blue,
}
 
\usepackage{cleveref}

\begin{document}

\title{Effects of finite volume and magnetic fields  on thermodynamic properties of quark matter and fluctuations of conserved charges} 
 \author{Nisha Chahal}
\email{nishachahal137@gmail.com}
\affiliation{Department of Physics, Dr. B R Ambedkar National Institute of Technology Jalandhar, 
 Jalandhar -- 144011, Punjab, India}
 
 \author{Suneel Dutt}
\email{dutts@nitj.ac.in}
\affiliation{Department of Physics, Dr. B R Ambedkar National Institute of Technology Jalandhar, 
 Jalandhar -- 144011, Punjab, India}
 
\author{Arvind Kumar}
\email{iitd.arvind@gmail.com, kumara@nitj.ac.in}
\affiliation{Department of Physics, Dr. B R Ambedkar National Institute of Technology Jalandhar, 
 Jalandhar -- 144011, Punjab, India}

\def\be{\begin{equation}}
\def\ee{\end{equation}}
\def\bearr{\begin{eqnarray}}
\def\eearr{\end{eqnarray}}
\def\zbf#1{{\bf {#1}}}
\def\bfm#1{\mbox{\boldmath $#1$}}
\def\hf{\frac{1}{2}}
\def\kp{\zbf k+\frac{\zbf q}{2}}
\def\km{-\zbf k+\frac{\zbf q}{2}}
\def\hwo{\hat\omega_1}
\def\hwt{\hat\omega_2}

\begin{abstract}
 In the current work, we present the influence of finite volume and magnetic field on the thermodynamic properties of isospin asymmetric quark matter using the Polyakov loop extended chiral SU(3) quark mean field (PCQMF) model at finite chemical potential and temperature. Within the PCQMF model, we use the scalar and vector field values in mean-field approximation to obtain the thermodynamic properties: pressure density, entropy density and energy density. The susceptibilities of conserved charges for strongly interacting matter for different system sizes as well as for different values of the magnetic field have been studied. A sizable shift in phase boundary towards the higher values of quark chemical potential ($\mu_q$) and temperature (T) has been observed for decreasing values of system volume as well as an opposite shift towards lower temperature and quark chemical potential for increasing magnetic field. We observe an enhancement in fluctuations of conserved charges in the regime of the transition temperature. These studies may have a significant role in understanding the thermodynamic observables extracted from heavy-ion collisions data.
\end{abstract}

\maketitle

\maketitle

\section{Introduction}
\label{intro}
The main aim of heavy-ion collision experiments across the world is to study the phase structure of the strongly interacting matter at extreme conditions of density and temperature \cite{busza2018heavy,ohnishi2012phase,pandav2022search}. Ongoing experimental facilities at Relativistic Heavy Ion Collider (RHIC) \cite{kumar2013star,odyniec2010rhic}, the Large Hadron Collider (LHC) \cite{bruning2012large}, as well as upcoming programs like Nuclotron-based Ion Collider facility (NICA) at JINR \cite{sissakian2009nuclotron} and
 Facility for Antiproton and Ion Research (FAIR) \cite{durante2019all} at GSI are aimed at exploring the different regions of quantum chromodynamics (QCD) phase diagram \cite{stephanov2006qcd}.

Alongside the experimental analysis, some hadron properties can also
be studied by numerical Lattice QCD simulations, which are a
non-perturbative application of field theory based on Feynman path integral approach \cite{detar2009qcd,gupta1999lattice}. These calculations, performed on various lattice points in a space-time grid have predicted a smooth crossover from confined hadronic state to deconfined quark-gluon plasma phase at high temperatures
and low baryonic densities \cite{aoki2006order,guenther2021overview}. At finite chemical potential values, lattice simulations face the challenge of the fermion sign problem \cite{goy2017sign,muroya2003lattice,danzer2009study}.
Other theoretical approaches based on non-perturbative QCD has shown the existence of first-order phase transition at low values of temperature and high baryonic density \cite{fodor2002new,ejiri2008canonical,hatta2003universality,fodor2004critical}. These two transition regimes are anticipated to be connected by a QCD critical point where the transition is expected to be of the second-order \cite{stephanov2004qcd}. The existence of the critical point, at a temperature value of about
165 MeV and baryonic chemical potential ($\mu_B$) value about 95 MeV, has been
suggested by Finite-Size Scaling (FSS) analysis of the data of the
Beam Energy Scan (BES) program at RHIC \cite{lacey2015indications}. In order to build upon the results of BES-I, BES Phase-II was
launched in 2019 to investigate the phase diagram in
intermediate-to-high baryonic chemical potential regimes with high
precision \cite{tlusty2018rhic}.

Determination of QCD properties like the equation of state and correlation functions in the neighborhood of a critical point is quite difficult, thus understanding the nature and precise measurement of the location of this point is a matter of ongoing high-priority research both experimentally and theoretically \cite{stephanov2005qcd}. Many theoretical models and approaches, such as hadron resonance gas (HRG)
model \cite{garg2013conserved}, Dyson–Schwinger  equation  framework \cite{maris2003dyson}, chiral hadronic model \cite{papazoglou1998chiral},
Nambu –Jona –Lasinio (NJL) model \cite{bowler1995nonlocal},  Polyakov-quark-meson  (PQM) model \cite{schaefer2007phase},  Polyakov  NJL (PNJL) model \cite{kashiwa2008critical}, and functional renormalization group (FRG) \cite{dupuis2021nonperturbative} approach has been employed to study the thermodynamic properties of the matter created in heavy-ion collisions.

The influence of non-zero magnetic field on the phase structure of QCD has attracted a great deal of attention as strong magnetic fields are believed to have been produced in the beginning of the universe \cite{harrison1973origin}, in magnetars \cite{enqvist1993primordial} as well as during heavy-ion collisions \cite{inghirami2020magnetic}. Thus it becomes essential to study the phase diagram at finite values of magnetic field along with non-zero chemical potential \cite{kharzeev2008effects,deng2012event}. During the electroweak transition in the early universe, the magnitude of the magnetic field is expected to be as high as $eB$ around 200$m_{\pi}^2$ \cite{vachaspati1991magnetic} while the value is estimated to be $eB$ around 0.1 $m_\pi^2$ for the CERN Super Proton Synchrotron (SPS) \cite{vallgren2011amorphous}, $eB$ of the order of $m_\pi^2$ at RHIC \cite{fu2013fluctuations}  and $eB$ approximately 15 $m_\pi^2$ at the CERN Large Hadron Collider \cite{skokov2009estimate}. In such heavy-ion collisions, charge separation is observed along the orbital momentum axis of the system, also known as the \enquote{Chiral Magnetic Effect} due to the
 strong magnetic fields generated in these collisions \cite{abelev2009azimuthal,kharzeev2008effects,fukushima2010chiral}. 
  The influence of external magnetic field on quark matter is determined by analyzing the interaction of valence and sea quarks \cite{balinew2012qcd}. In the vicinity of the critical temperature, quark condensate is enhanced due to the valence quarks, but its effect is saturated by sea quarks suppression \cite{bruckmann2013inverse}. Hence we observe a relocation of pseudo-critical temperature to lower values with the increase in the magnetic field, termed \enquote{inverse magnetic catalysis}. Seemingly contradictory results were obtained in early studies of low energy effective theories and lattice QCD simulations at vanishing values of temperature showing the enhancement of quark condensates, referred as \enquote{magnetic catalysis} \cite{shushpanov1997quark,klimenko1992three,shovkovy2013magnetic,bali2013thermodynamic,klevansky1989chiral}. Magnetic catalysis at zero temperature is attributed to the positive behavior of $\beta$ function in scalar QED \cite{endrHodi2013qcd}. The lattice calculations at more finer grid points and consideration of physical quark masses have also shown the relocation of critical point to a lower temperature value \cite{bali2012qcd}. The effect of the external magnetic field in the PNJL model has been shown to be the catalyzer of dynamical symmetry breaking in quark matter.

In the initial time period after a heavy-ion collision, a finite extent fireball is created, which extends up to the range of a few femtometers in all directions. Also, the possible critical-point and its impact is critically blurred out, considering the brief lifetime and small volume of QGP created in heavy-ion collisions \cite{fraga2011finite}. Thus the impact of finite volume on the thermodynamics of matter created after the collision as well as phase transition is quite significant \cite{bzdak2020mapping,luo2017search}. The motivation behind finite volume considerations in studying strong interactions can also be highlighted by FSS analysis \cite{bhattacharyya2017polyakov,fisher1972scaling}, which is a potent statistical tool working on the 
 idea of comparing the linear dimension L to correlation length, $\xi$ \cite{brezin1985finite}.
Theoretical studies have shown the repositioning of critical temperature and chemical potential towards higher values with the decrease in the system volume \cite{magdy2019influence,magdy2017influence}.
The impact of finite size has also been studied by considering spherical and cubic regions in the NJL model \cite{mata2022effects}. Finite size effects have been incorporated in the PNJL model by employing Multiple Reflection
Expansion (MRE) formalism \cite{grunfeld2018finite,kiriyama2005color}. This framework describes a sphere rather than a cube and is included in the modification of the density of states
\cite{zhao2019chiral,zhao2020finite,lugones2019surface}.

Fluctuations and correlations of conserved charges, which are measurable by event-by-event analysis of heavy-ion collisions, carries important information regarding the hot matter created in the collision \cite{borsanyi2012fluctuations,adam2020net,chahal2022quark,chatterjee2012fluctuations,fan2019probing}. These have been identified as the observables that can be calculated both experimentally and in theoretical models to have a more precise knowledge of the critical end-point (CEP). Most importantly, the non-Gaussian behavior of susceptibilities of conserved charges has recently grabbed much of attention \cite{asakawa2016fluctuations}. In lattice QCD, the Polyakov loop susceptibilities in the finite size have been estimated by choosing lattice sizes in the range of palpable volume realized in  heavy ion collisions \cite{berg20133}. Using a similar approach, quark number susceptibilities and thermodynamic properties have been investigated by applying Monte-Carlo simulations to PNJL model \cite{cristoforetti2010thermodynamics}. It has been shown in \cite{ding2021fluctuations} that with increasing value of finite size and magnetic field, susceptibilities of conserved quantities have shown significant enhancement of the peaked structure near the transition regime in the QCD phase diagram. The sudden rise in the skewness of the baryon number and strangeness number is highlighted as an essential parameter to explore the QCD critical-end point \cite{fu2013fluctuations}.

In the present work, we aim to inspect the thermodynamic properties and fluctuations of conserved charges in quark matter, considering the impact of finite volume and finite value of the magnetic field. For this, we have used the formalism of Polyakov loop extended chiral SU(3) quark mean field model \cite{wang2003strange}. In order to describe a complex many-body problem, we need to consider a relativistic quantum field theory that contains scalar and vector meson fields alongside baryons. In the well-studied phenomenology of a nucleon-nucleon interactions, attractive contribution has been postulated at an intermediate range while the repulsive contribution is populated at a shorter range. These repulsive interactions are modeled in mean-field model by vector mesons, whereas scalar mesons are included to define the attractive interactions \cite{poberezhnyuk2017quantum}. This model comprehends the interaction of quarks by the inclusion of spin-0 and even parity scalar meson fields along with spin-1 and odd parity vector meson fields. The scalar fields are non-strange scalar field ($\sigma$), strange scalar field ($\zeta$), scalar isovector field ($\delta$) and isoscalar dilaton field ($\chi$). The vector fields included in the model are non-strange vector field ($\omega$), vector-isovector field ($\rho$), and strange vector field ($\phi$). This model has been used in literature to study the properties of finite nuclei \cite{wang2002multi}, hypernuclei, \cite{wang2004new} and strange hadronic matter \cite{wang2001strange}. The Polyakov loop potential is included in the chiral SU(3) quark mean field model (CQMF) in order to give a better description of the deconfinement transition and to incorporate the thermal fluctuations arising in the pure gluonic theory \cite{fukushima2017polyakov,chahal2022quark}. With the introduction of the Polyakov loop, theoretical models give a better reproduction of lattice data \cite{hansen2020quark,ratti2006phases}. Quantum and thermal fluctuations of the matter can also be included in theoretical models by using the functional renormalization group (FRG) approach, which is based on flow equations derived from effective action \cite{herbst2011phase,fu2021high}. 

This paper is organized as follows: In Sect.~\ref{method}, we have briefly described the PCQMF and have obtained the grand canonical potential and thermodynamic quantities. In Sect.~\ref{taylor}, Taylor series method is discussed in detail for the description of cumulants and hence susceptibilities. In Sect.~\ref{results}, a descriptive view of all the results for thermodynamic properties of quark matter, a phase diagram of QCD for finite volume and magnetic field and fluctuations has been presented. In Sect.~\ref{summary}, we have summarised the results of the present work. 

\section{Methodology}
\subsection{Polyakov chiral  SU(3) quark mean field model} \label{method}
CQMF model incorporates the meson-meson and quark-meson interactions and is based on chiral $SU(3)_L \times SU(3)_R$ symmetry, it's spontaneous breaking \cite{bando1988nonlinear,weinberg1968nonlinear}, and also the broken scale invariance \cite{kharzeev2009broken}. In this model, quarks are bound in the hadrons by an effective mean field potential and are used to explain many-body interactions based on a relativistic non-perturbative approach. The masses of pseudoscalar mesons are derived from explicit symmetry breaking, while the masses of quarks and other mesons are attributed to the spontaneous symmetry breaking \cite{beekman2019introduction}. The total effective Lagrangian density of the model in order to study strange quark matter is written as \cite{kumari2021quark}
 \begin{equation}\label{eqn1}
     \mathcal{L}_{\text{eff}} = \mathcal{L}_{\text {q0}}+\mathcal{L}_{\text {qm}}+
\mathcal{L}_ {\Sigma\Sigma} + \mathcal{L}_{\text {VV}} 
+\mathcal{L}_{S B}+\mathcal{L}_{\Delta m} + \mathcal{L}_{\text {h}},
 \end{equation}
where  ${\cal L}_{q0} = \bar q \, i\gamma^\mu \partial_\mu q $ gives the free part of massless quarks. The quark meson interaction term, ${\cal L}_{qm}$ splits into right and left-handed parts in chiral limits and is represented as \cite{chahal2022quark}

\begin{equation}\label{eqn2}
 	\mathcal{L}_{q m}=g_{s}\left(\bar{\psi}_{L} M \psi_{R}+\bar{\psi}_{R} M^{\dagger} \psi_{L}\right)-g_{v}\left(\bar{\psi}_{L} \gamma^{\mu} l_{\mu} \psi_{L}+\bar{\psi}_{R} \gamma^{\mu} r_{\mu} \psi_{R}\right).
 \end{equation}

In the above equation, the quark spinor, $\psi$ = $(u,d,s)$  for $N_c$ = 3, color degrees of freedom, $g_v$ and $g_s$ define the coupling strength of vector and scalar mesons with quarks, respectively. The nonets for spin-zero and one mesons can be written as
\begin{equation} \label{eqn3}
 M\left(M^{\dagger}\right)=\Sigma \pm i \Pi=\frac{1}{\sqrt{2}} \sum_{a=0}^{8}\left(\sigma^{a} \pm i \pi^{a}\right) \lambda^{a}
\end{equation}

and,

 \begin{equation} \label{eqn4}
     l_{\mu}\left(r_{\mu}\right)=\frac{1}{2}\left(V_{\mu} \pm A_{\mu}\right)=\frac{1}{2 \sqrt{2}} \sum_{a=0}^{8}\left(v_{\mu}^{a} \pm a_{\mu}^{a}\right) \lambda^{a}.
\end{equation}

In equation~\ref{eqn3}, $\Sigma$ and $\Pi$ represent the spin-0 scalar and pseudoscalar mesons. Nonets of vector and pseudovector mesons are given by $v_{\mu}^a$ and $a_{\mu}^a$, respectively. The third and the fourth term in equation~\ref{eqn1} expresses the self-interactions of scalar mesons and vector mesons, respectively. The attractive contribution of interactions in the medium is attributed by the inclusion of scalar mesons. These self-interaction terms of mesons are given as

\begin{eqnarray}
{\cal L}_{\Sigma\Sigma} =& -\frac{1}{2} \, k_0\chi^2
\left(\sigma^2+\zeta^2+\delta^2\right)+k_1 \left(\sigma^2+\zeta^2+\delta^2\right)^2
+k_2\left(\frac{\sigma^4}{2} +\frac{\delta^4}{2}+3\sigma^2\delta^2+\zeta^4\right)\nonumber \\ 
&+k_3\chi\left(\sigma^2-\delta^2\right)\zeta 
 -k_4\chi^4-\frac14\chi^4 {\rm ln}\frac{\chi^4}{\chi_0^4} +
\frac{d}
3\chi^4 {\rm ln}\left(\left(\frac{\left(\sigma^2-\delta^2\right)\zeta}{\sigma_0^2\zeta_0}\right)\left(\frac{\chi^3}{\chi_0^3}\right)\right), \label{scalar0}
\end{eqnarray}

\begin{equation}
\mathcal{L}_{V V}=\frac{1}{2} \frac{\chi^{2}}{\chi_{0}^{2}}\left(m_{\omega}^{2} \omega^{2}+m_{\rho}^{2} \rho^{2}+m_{\phi}^{2} \phi^{2}\right)+g_{4}\left(\omega^{4}+6 \omega^{2} \rho^{2}+\rho^{4}+2 \phi^{4}\right).
\end{equation}
The interaction of mesons with quarks is described by the interchange of scalar as well as vector meson fields. In order to study strange matter, it is quite important to include a strange scalar isoscalar field, $\zeta$, due to its strange quark content. The scalar isovector field, $\delta$, gives the description of isospin asymmetric matter. In reality, chiral SU(3) symmetry is not satisfied exactly by quark-meson interactions. This is due to the fact that pseudoscalar mesons are Goldstone bosons with zero mass, but the masses of $K$ and $\pi$ mesons are not zero. Hence, the contribution of explicit symmetry breaking in the effective Lagrangian is included by ${\cal L}_{SB}$, ${\cal L}_{\Delta m}$ and ${\cal L}_{h}$ \cite{wang2002multi,papazoglou1999nuclei}. The term ${\cal L}_{SB}$ gives rise to non-zero masses for pseudoscalar mesons and is given by

\begin{equation} \label{lsb}
{\cal L}_{SB}=-\frac{\chi^2}{\chi_0^2}\left[m_\pi^2f_\pi\sigma + 
\left(\sqrt{2}m_K^2f_K-\frac{m_\pi^2}{\sqrt{2}}f_\pi\right)\zeta\right].
\end{equation}

The partially conserved axial-vector current (PCAC) relations are thus satisfied for $K$ and $\pi$ mesons as a result of non-vanishing divergence of axial currents. An additional mass term is included in the model, which gives the reasonable constituent mass of strange quark, represented as
\begin{eqnarray} \label{lm}
{\cal L}_{\Delta m} = - \Delta m_s \bar q S q,
\end{eqnarray}
where $S \, = \, \frac{1}{3} \, \left(I - \lambda_8\sqrt{3}\right)$ is the strangeness quark matrix. The term ${\cal L}_h \, = \, (h_1 \, \sigma \,  + \, h_2 \, \zeta) \, \bar{s} s \ $, describes hyperon potential in the mean-field approximation. 

For good reproduction of lattice data, effective chiral models are extended by the introduction of different forms of Polyakov-loop potential. Due to the success of this approach, eﬀective models can be seen as an enticing method to study strange hadronic and quark matter. Hence it becomes possible to study both chiral symmetry breaking and deconfinement in the upgraded model. Thus the total Lagrangian of the PCQMF model is defined as
\begin{equation}
{\cal L}_{{\rm PCQMF}} \, = \, {\cal L}_{\rm eff} \, -\mathcal{U}(\Phi(\vec{x}),\bar{\Phi}(\vec{x}),T), \label{PCQMFlag}
\end{equation}
where $\mathcal{U}(\Phi(\vec{x}),\bar{\Phi}(\vec{x}),T)$  is the effective Polyakov loop potential. In the above equation, $\Phi$ and $\bar{\Phi}$ are the Polyakov-loop variables and are defined as the expectation value of trace over the color of the thermal Wilson line \cite{kumari2021quark}. Polyakov loop potential is not uniquely defined and can be constructed from the center symmetry of the pure-gauge theory. For this work, we consider the polynomial parameterized form of Polyakov loop potential, which is given by \cite{PhysRevD.81.074013,PhysRevD.75.034007}

\begin{equation}
\frac{\mathcal{U}_{\text {poly }}(\Phi, \bar{\Phi})}{T^{4}}=-\frac{b_{2}(T)}{2} \bar{\Phi} \Phi-\frac{b_{3}}{6}\left(\Phi^{3}+\bar{\Phi}^{3}\right)\\
+\frac{b_{4}}{4}(\bar{\Phi} \Phi)^{2}.
\end{equation}
The temperature-dependent coefficient, $b_{2}(T)$, appearing in the above equation, is defined as
\begin{equation}
   b_{2}(T)=a_{0}+a_{1}\left(\frac{T_{0}}{T}\right)+a_{2}\left(\frac{T_{0}}{T}\right)^{2}+a_{3}\left(\frac{T_{0}}{T}\right)^{3} . 
\end{equation}
The parameters are determined by fitting the lattice simulation data and hence we consider the following: $a_0=1.53$, $a_1=0.96$, $a_2=-2.3$, $a_3=-2.85$, $b_3=13.34$ and $b_4=14.88$. Here $T_0$ is the critical temperature for the change of phase from confined hadrons to deconfined quarks in the pure Yang-Mills theory at a very low value of chemical potential \cite{fukugita1990finite}. Thus the thermodynamic potential for the PCQMF model is defined as 
 \begin{equation}\label{omegapnjl}
\Omega=\mathcal{U}(\Phi,\bar{\Phi},T)
+\Omega_{q\bar{q}}-{\cal L}_M- {\cal V}_{vac}.
\end{equation}

In the above equation, ${\cal L}_M = {\cal L}_{VV}+ \mathcal{L}_{\Sigma\Sigma} + {\cal L}_{SB}$, is the meson interaction term. The vacuum term, ${\cal V}_{vac}$ is subtracted in order to obtain zero vacuum energy. 
The thermal contribution of quarks and antiquarks to the total thermodynamic potential is represented by 
 \begin{eqnarray}\label{omegaq}
\Omega_{q\bar{q}}=-\sum_{i=u,d,s} \gamma_i T \int_0^\infty\frac{d^3p}{(2\pi)^3}[\ln(g_i^{+})+\ln(g_i^{-})].
\end{eqnarray}
In the above equation, $g_{i}^{+}$ and $g_{i}^{-}$ are defined as 
 \begin{equation}
 g_{i}^{+}=\left[1+3 \Phi e^{-(E_i^{*}-\nu_i^{*})/ T}+3 \bar{\Phi} e^{-2(E_i^{*}-\nu_i^{*}) / T}+e^{-3(E_i^{*}-\nu_i^{*})/ T}\right],\\ 
 \end{equation}
 and
 \begin{equation}
 g_{i}^{-}=\left[1+3\bar\Phi e^{-(E_i^{*}+\nu_i^{*})/ T} +3 \Phi e^{-2(E_i^{*}+\nu_i^{*})/ T}  +e^{-3(E_i^{*}+\nu_i^{*})/ T}\right].
 \end{equation}
Here the summation is over the constituent quarks, and $\gamma_i$ is the spin degeneracy factor. In the above equation, $\nu_{i}^{*}=\mu_{i}-g_{\omega i} \omega-g_{\phi i} \phi-g_{\rho i} \rho$ is the effective chemical potential derived from vector fields. The effective single particle energy is calculated in terms of the effective constituent mass of quarks, 
${m_i}^{*} = -g_{\sigma i}\sigma - g_{\zeta i}\zeta - g_{\delta i}\delta + {m_{i0}}$, where $m_{u0} = m_{d0}$ = 0 and $m_{s0}$ = 29 MeV. The $g_{\sigma i}$, $g_{\zeta i}$, $g_{\delta i}$ give the coupling strength of different quarks with scalar mesons and $g_{\omega i}$,  $g_{\rho i}$, $g_{\phi i}$ for vector mesons. The effective energy of quarks is defined as  
\begin{equation}
E_i^{*}=\sqrt{p^2 + m_{i}^{*2}}.
\end{equation}

Now, in order to study the impact of magnetic field on the strange quark matter in PCQMF model, we consider a homogeneous magnetic field, B in the $z$-direction. In the presence of the finite magnetic field, the total effective energy of the quarks is modified as \cite{tawfik20143,ferreira2014inverse}
\begin{equation}
 E_{i}^{*}=\sqrt{p_{z}^{2}+m_{i}^{*2}+\left|q_{i}\right|(2 n+1-\Upsilon) B},   
\end{equation}
and
\begin{equation}
    p = \sqrt{p_z^{2}+\left|q_{i}\right|(2 n+1-\Upsilon) B}.
\end{equation}
The above equation gives the relation of total momentum, $p$ with the longitudinal momentum, $p_z$ and $\Upsilon$ defines the spin quantum number S($\Upsilon = \pm S/2$). Now, $2n + 1 - \Upsilon$ can be replaced by a single quantum number, $k$, known as Landau level. Thus the total thermodynamical potential is altered, and the term giving the contribution of quarks and antiquarks interaction is written as \cite{fu2013fluctuations,magdy2017influence}

\begin{equation} \label{mag}
\Omega_{q \bar q} = -\sum_{i=u, d, s} \frac{\left|q_{i}\right| B T}{2 \pi} \sum_{k=0}^{\infty} \alpha_{k} \int_{-\infty}^{\infty} \frac{d p_{z}}{2 \pi}\left( \ln { g_{i}^{+} }+ \ln {g_{i}^{-}} \right).
\end{equation}

The summation in equation~\ref{mag} runs over the lowest landau level, $k$ = 0 to the maximum occupied Landau level \cite{tawfik20183}. Here $\alpha_k$ is the spin-degeneracy factor and due to the polarization of charged particles in the lowest Landau levels by the external magnetic field, its value is 1 for $k$ = 0 and 2 otherwise. The impact of the finite size effect is assimilated in the model by using the approximation method defined in \cite{bhattacharyya2013thermodynamic,bhattacharyya2015thermodynamics,magdy2019influence} by introducing a lower momentum cutoff, $p_{min}$ [MeV] = $\pi/R$ [MeV] = $\Lambda$, where $R$ is the length of a cubic volume. In order to consider the complete execution of finite volume, one has to carefully analyze the impact of the surface and curvature of the system under examination. Along with this, periodic boundary conditions for bosons and  anti-periodic conditions for fermions have to be scrutinized, leading to the indeterminable sum over discrete momentum values \cite{magdy2019influence}. In NJL and PNJL models, the fermion vacuum term is accountable for dynamical chiral symmetry breaking in the vacuum and is thus considered with proper divergence regularization of a ultraviolet cutoff parameter \cite{skokov2010vacuum}. On the contrary, spontaneous breaking of chiral symmetry is incorporated through the mesonic
potential in the current model, as in PQM and PLSM models \cite{mao2010phase,schaefer2007phase}. Due to the inclusion of the fermion vacuum term in the PQM model, the critical temperature has been reported to shift towards the higher values of baryonic chemical potential and low values of temperature \cite{chatterjee2012including,gupta2012revisiting}. The fermion vacuum term has not been considered in the current work and hence, no upper cutoff on momentum has been implied. 

In order to calculate the values of different fields at the varying value of temperatures and density, we minimize the total thermodynamic potential obtained after the inclusion of external magnetic field and finite volume. Thus we have,
\begin{equation}
\frac{\partial\Omega}
{\partial\sigma}=\frac{\partial\Omega}{\partial\zeta}=\frac{\partial\Omega}{\partial\delta}=\frac{\partial\Omega}{\partial\chi}=\frac{\partial\Omega}{\partial\omega}=\frac{\partial\Omega}{\partial\rho}=\frac{\partial\Omega}{\partial\phi}=\frac{\partial\Omega}{\partial\Phi}=\frac{\partial\Omega}{\partial\bar\Phi}=0.
\end{equation}
Plugging $\Omega$ in above, following system of equations is obtained:
\begin{eqnarray}\label{sigma1}
&&\frac{\partial \Omega}{\partial \sigma}= k_{0}\chi^{2}\sigma-4k_{1}\left( \sigma^{2}+\zeta^{2}
+\delta^{2}\right)\sigma-2k_{2}\left( \sigma^{3}+3\sigma\delta^{2}\right)
-2k_{3}\chi\sigma\zeta \nonumber\\
&-&\frac{d}{3} \chi^{4} \bigg (\frac{2\sigma}{\sigma^{2}-\delta^{2}}\bigg )
+\left( \frac{\chi}{\chi_{0}}\right) ^{2}m_{\pi}^{2}f_{\pi}- 
\left(\frac{\chi}{\chi_0}\right)^2m_\omega\omega^2
\frac{\partial m_\omega}{\partial\sigma}\nonumber\\
 &-&\left(\frac{\chi}{\chi_0}\right)^2m_\rho\rho^2 
\frac{\partial m_\rho}{\partial\sigma}
-\sum_{i=u,d} g_{\sigma i}\rho_{s i} = 0 ,
\end{eqnarray}
\begin{eqnarray}
&&\frac{\partial \Omega}{\partial \zeta}= k_{0}\chi^{2}\zeta-4k_{1}\left( \sigma^{2}+\zeta^{2}+\delta^{2}\right)
\zeta-4k_{2}\zeta^{3}-k_{3}\chi\left( \sigma^{2}-\delta^{2}\right)-\frac{d}{3}\frac{\chi^{4}}{{\zeta}}\nonumber\\
&+&\left(\frac{\chi}{\chi_{0}} \right)
^{2}\left[ \sqrt{2}m_{K}^{2}f_{K}-\frac{1}{\sqrt{2}} m_{\pi}^{2}f_{\pi}\right]-\left(\frac{\chi}{\chi_0}\right)^2m_\phi\phi^2 
\frac{\partial m_\phi}{\partial\zeta}
 -\sum_{i=s} g_{\zeta i}\rho_{s i} = 0 ,
\label{zeta}
\end{eqnarray}
\begin{eqnarray}
\frac{\partial \Omega}{\partial \delta}=k_{0}\chi^{2}\delta-4k_{1}\left( \sigma^{2}+\zeta^{2}+\delta^{2}\right)
\delta-2k_{2}\left( \delta^{3}+3\sigma^{2}\delta\right) +\mathrm{2k_{3}\chi\delta
\zeta} \nonumber\\
 +  \frac{2}{3} d \chi^4 \left( \frac{\delta}{\sigma^{2}-\delta^{2}}\right)
-\sum_{i=u,d} g_{\delta i}\rho_{s i} = 0 ,
\label{delta}
\end{eqnarray}
\begin{eqnarray}
&&\frac{\partial \Omega}{\partial \chi}=\mathrm{k_{0}\chi} \left( \sigma^{2}+\zeta^{2}+\delta^{2}\right)-k_{3}
\left( \sigma^{2}-\delta^{2}\right)\zeta + \chi^{3}\left[1
+{\rm {ln}}\left( \frac{\chi^{4}}{\chi_{0}^{4}}\right)  \right]
+(4k_{4}-d)\chi^{3}
\nonumber\\
&-&\frac{4}{3} d \chi^{3} {\rm {ln}} \Bigg ( \bigg (\frac{\left( \sigma^{2}
-\delta^{2}\right) \zeta}{\sigma_{0}^{2}\zeta_{0}} \bigg )
\bigg (\frac{\chi}{\mathrm{\chi_0}}\bigg)^3 \Bigg )+
\frac{2\chi}{\chi_{0}^{2}}\left[ m_{\pi}^{2}
f_{\pi}\sigma +\left(\sqrt{2}m_{K}^{2}f_{K}-\frac{1}{\sqrt{2}}
m_{\pi}^{2}f_{\pi} \right) \zeta\right] \nonumber\\
&-& \frac{\chi}{{\chi^2}_0}({m_{\omega}}^2 \omega^2+{m_{\rho}}^2\rho^2)  = 0,
\label{chi}
\end{eqnarray}
  \begin{eqnarray}
\frac{\partial \Omega}{\partial \omega}=\frac{\chi^2}{\chi_0^2}m_\omega^2\omega+4g_4\omega^3+12g_4\omega\rho^2
&-&\sum_{i=u,d}g_{\omega i}\rho_{v i}=0,
\label{omega} 
\end{eqnarray}
  \begin{eqnarray}
\frac{\partial \Omega}{\partial \rho}=\frac{\chi^2}{\chi_0^2}m_\rho^2\rho+4g_4\rho^3+12g_4\omega^2\rho&-&
\sum_{i=u,d}g_{\rho i}\rho_{v i}=0, 
\label{rho} 
\end{eqnarray}
  \begin{eqnarray}
\frac{\partial \Omega}{\partial \phi}=\frac{\chi^2}{\chi_0^2}m_\phi^2\phi+8g_4\phi^3&-&
\sum_{i=s}g_{\phi i}\rho_{v i}=0,
 \label{phi}  
\end{eqnarray}
 \begin{eqnarray}
\hspace*{0.4cm} 
\frac{\partial \Omega}{\partial \Phi} =\bigg[\frac{-a(T)\bar{\Phi}}{2}-\frac{6b(T)
(\bar{\Phi}-2{\Phi}^2+{\bar{\Phi}}^2\Phi)
}{1-6\bar{\Phi}\Phi+4(\bar{\Phi}^3+\Phi^3)-3(\bar{\Phi}\Phi)^2}\bigg]T^4
-\sum_{i=u,d,s}\frac{2TN_C}{(2\pi)^3}
\nonumber\\
\int_0^\infty d^3k 
\bigg[\frac{e^{-(E_i^*-{\nu_i}^{*})/T}}{g_i^{+}}
+\frac{e^{-2(E_i^*+{\nu_i}^{*})/T}}{g_i^{-}}\bigg]=0,
\label{Polyakov} 
\end{eqnarray}
and
  \begin{eqnarray}
\frac{\partial \Omega}{\partial \bar{\Phi}} =\bigg[\frac{-a(T)\Phi}{2}-\frac{6b(T)
(\Phi-2{\bar{\Phi}}^2+{\Phi}^2\bar{\Phi})
}{\mathrm{1-6\bar{\Phi}\Phi+4(\bar{\Phi}^3+\Phi^3)-3(\bar{\Phi}\Phi)^2}}\bigg]T^4
-\sum_{i=u,d,s}\frac{2TN_C}{(2\pi)^3}
\nonumber\\
\int_0^\infty d^3k\ \bigg[\frac{e^{-2(E_i^*-{\nu_i}^{*})/T}}{g_i^{+}}
+\frac{e^{-(E_i^*+{\nu_i}^{*})/T}}{g_i^{-}}\bigg]=0. 
\label{Polyakov conjugate} 
\end{eqnarray}

The coupling constants $g_{\sigma i}$, $g_{\zeta i}$ and $m_{s0}$ are calculated by fitting the vacuum masses of constituent quarks as $m_u = m_d$ = 313 MeV and $ms$ = 490 MeV. The free model parameters $h_1$, $h_2$, $k_0$, $k_1$, $k_2$, $k_3$, $k_4$, $g_s$ and $g_v$ are calculated by considering the vacuum masses of $\sigma$, $\zeta$, and $\chi$ meson along with the masses of $K$ and $\pi$ mesons and mean masses of $\eta$ and ${\eta}^{'}$ mesons which are expressed by the eigenvalues of the mass matrix \cite{ping2001nuclear}. The relations between various quark meson coupling constants is given as

\begin{equation}
    \frac{g_s}{\sqrt{2}} = g_\sigma^u=g_\sigma^d=\frac{1}{\sqrt{2}} g_\zeta^s=g_\delta^u=-g_\delta^d, \quad g_\sigma^s=g_\zeta^u=g_\zeta^d=g_\delta^s=0 
    \end{equation}

\begin{equation}
\frac{g_v}{2 \sqrt{2}}=g_\omega^u=g_\omega^d=g_{\rho}^{u}=-g_{\rho}^d=\frac{1}{\sqrt{2}} g_\phi^s, \quad g_\omega^s=g_{\rho}^s=g_\phi^u=g_\phi^d=0
\end{equation}

The flavor asymmetry is assimilated in the model by including isospin ($\mu_I$) and strangeness chemical potential, $\mu_S$. The baryon, strangeness, and isospin chemical potential are determined through relations

\begin{equation}
\begin{array}{l}

    \mu_B = \frac{3}{2}(\mu_u + \mu_d), \\
    \mu_S = \frac{1}{2}(\mu_u + \mu_d - 2\mu_s), \\
    \mu_I = \frac{1}{2}(\mu_u - \mu_d).
    \end{array}
\end{equation}

The vector and scalar density of quarks are defined as
\begin{eqnarray}
\rho_{vi} = \sum_{i=u, d, s} \frac{ N_c \left |q_{i}\right| B }{2 \pi} \sum_{k=0}^{\infty} \alpha_{k} \int_{-\infty}^{\infty} \frac{d p_{z}}{2 \pi}\left( f_i+\bar{f}_i \right), 
\label{rhov}
\end{eqnarray}
 and
\begin{eqnarray}
\rho_{si} = \sum_{i=u, d, s} \frac{ N_c \left |q_{i}\right| B }{2 \pi} \sum_{k=0}^{\infty} \alpha_{k} \int_{-\infty}^{\infty} \frac{d p_{z}}{2 \pi} \frac{m_{i}^{*}}{E^{\ast}_i(k)} \left( f_i+\bar{f}_i \right). 
\label{rhos}
\end{eqnarray}

In the above equations, $f_i$, and $\bar{f}_i$ are the Fermi distribution functions at the finite value of temperature for quarks and anti-quarks and are defined as

\begin{equation}\label{distribution}
  f_{i}=\frac{\Phi e^{-(E_i^*-{\nu_i}^{*})/T}+2\bar{\Phi} e^{-2(E_i^*-{\nu_i}^{*})/T}+e^{-3(E_i^*-{\nu_i}^{*})/T}}
  {g_i^{+}} , 
\end{equation}

\begin{equation}\label{distribution1}
  \bar{f}_{i}=\frac{\bar{\Phi} e^{-(E_i^*+{\nu_i}^{*})/T}+2{\Phi} e^{-2(E_i^*+{\nu_i}^{*})/T}+e^
  {-3(E_i^*+{\nu_i}^{*})/kT}}{g_i^{-}} .
\end{equation}

\subsection{Thermodynamic quantities and Taylor series expansion} \label{taylor}
After the calculation of $\sigma$, $\zeta$, and $\delta$, the dilaton field $\chi$, the vector fields $\omega$, $\rho$ and $\phi$ and the Polyakov fields $\Phi$ and its conjugate $\bar{\Phi}$, the thermodynamic potential density is used to calculate the pressure, $p$, entropy density, $s$ and the energy density, $\epsilon$, given by
\begin{equation}
p=-\Omega,
\label{p1}
\end{equation}

\begin{equation}
s=-\frac{\partial \Omega}{\partial T},
\end{equation}
and
\begin{equation}
\epsilon=\Omega+\sum_{i=u, d, s} {\nu_i}^{*} \rho_i+TS.
\label{energy1}
\end{equation}

By substituting the value of pressure derived from the above equation, we can write a generalized expression for the susceptibilities of conserved charges as \cite{shao2018baryon}
\begin{equation}\label{sus}
    \chi^{BQS}_{ijk} =  \frac {\partial^{i+j+k} [P/T^{4}]} {\partial \left(\mu_{B}/T \right)^{i}
\partial(\mu_{Q}/T)^{j} \partial(\mu_{S}/T)^{k}},
\end{equation}

here, $\mu_B$ is the baryon chemical potential, $\mu_Q$ is the charge number chemical potential, and $\mu_S$ stands for the strangeness chemical potential. In heavy ion collision experiments, the susceptibilities of conserved charges in equation~\ref{sus} are also correlated to the cumulants of conserved charge multiplicity distributions. The susceptibilities and correlations are hence computed in terms of the ensemble average of these conserved quantities, 
$\delta N_X = N_X - \langle N_X \rangle$ \cite{asakawa2016fluctuations,cheng2009baryon}.



In order to have a better understanding of the experimentally derived observables, these fluctuations can be evaluated theoretically by different methods. In \cite{skokov2011charge}, the net quark number density fluctuations and higher-order cumulants have been investigated by using FRG approach in Polyakov quark meson model. Baryon and quark number fluctuations have also been calculated by the Dyson-Schwinger equation approach in \cite{isserstedt2020dyson,xin2014quark}. We use Taylor series expansion in the current work to compute the susceptibilities of conserved charges \cite{PhysRevD.73.114007,schmidt2010net}. The scaled pressure in equation~\ref{sus} is expanded for $\mu_{(B,Q,S)} = 0$ as

\begin{equation}\label{cumu}
    \frac{P(T,\mu_{(B,Q,S)})}{T^{4}} = \sum_{n=0}^{\infty} {c_{n}(T)}\left(\frac{\mu_{(B,Q,S)}}{T} \right)^n .
\end{equation}
So the coefficients of the above series, give the derivatives of conserved charges at vanishing value of chemical potential and thus the susceptibilities.

\section{Results and discussion}  \label{results}
In this section, we present the results for different thermodynamic properties of isospin asymmetric quark matter due to the inclusion of finite volume and magnetic field in the Polyakov loop extended chiral SU(3) quark mean-field model. The values of different fields at varying temperature and density values are obtained by solving the non-linear coupled equations mentioned earlier. The parameters used in the current study are listed in Table~\ref{table1} and Table~\ref{table2}. In section~\ref{thermosub}, we have discussed the in-medium behavior of scalar and vector fields and the Polyakov loop field in the presence of a finite magnetic field and finite volume. Additionally, in-medium masses of quarks and various thermodynamic quantities have also been studied. In section~\ref{sussub}, we have discussed in detail the impact of finite size and magnetic field on the susceptibilities of conserved charges.

\begin{table}[b]
\centering
\begin{tabular}{|c|c|c|c|c|c|c|c|c|}
\hline
 $\sigma_0$ (MeV) & $\zeta_0$(MeV)  & $\chi_0$(MeV)   & $m_\pi$(MeV)  & $f_\pi$(MeV)  & $m_K$(MeV)                     \\ \hline
   -93                  & -96.87              & 254.6               & 139                & 93                 & 496                                                                  \\ \hline
$f_K$(MeV)     & $m_\omega$(MeV)  & $m_\phi$(MeV)  & $m_\rho$( MeV) &$\rho_0$(fm$^{-3}$)  &   \\ \hline
115                 & 783  &  1020                & 783   &    0.15  &                                              \\ \hline
\end{tabular}
\caption{The list of constant values used to fit the model's parameters.}
\label{table1}
\end{table}

\begin{table}[b]
\centering
\begin{tabular}{|c|c|c|c|c|c|c|c|c|}
\hline
$k_0$           & $k_1$          & $k_2$          & $k_3$         & $k_4$         & $g_s$         & $\rm{g_v}$          & $\rm{g_4}$                                              \\ \hline
4.94                 & 2.12                & -10.16              & -5.38              & -0.06              & 4.76               & 1.95               & 37.5                                                            \\ \hline
$d$   & $g_{\zeta d}$ & $g_{\zeta s}$ & $g_{\delta u}$  & $g_{\delta d}$  & $g_{\delta s}$   &  $g_{\omega u}$  & $g_{\omega d}$                               \\ \hline
 0.18  & 0                  & 4.76               & 3.36  & -3.36                & 0                   &                    0.336                 & 0.336                                                             \\ \hline
$g_{\omega s}$ & $g_{\phi u}$  & $g_{\phi d}$  & $g_{\phi s}$  & $g_{\rho u}$   & $g_{\rho d}$    & $g_{\rho s}$   & $h_1$ \\ \hline
   0                   & 0                  & 0                  & 0.975  & 0.336                & -0.336                & 0                   & -2.20  \\ \hline
   $h_2$  &  $g_{\sigma u}$  & $g_{\sigma d}$ & $g_{\sigma s}$ & $g_{\zeta u}$ &   &    & \\ \hline
3.24           &  3.36                 & 3.36                & 0                   & 0    &   &  &                                      \\ \hline

\end{tabular}
\caption{The list of fitted parameters used in the current work.}
\label{table2}
\end{table}

\begin{figure}
    \centering
    \includegraphics[scale=0.605]{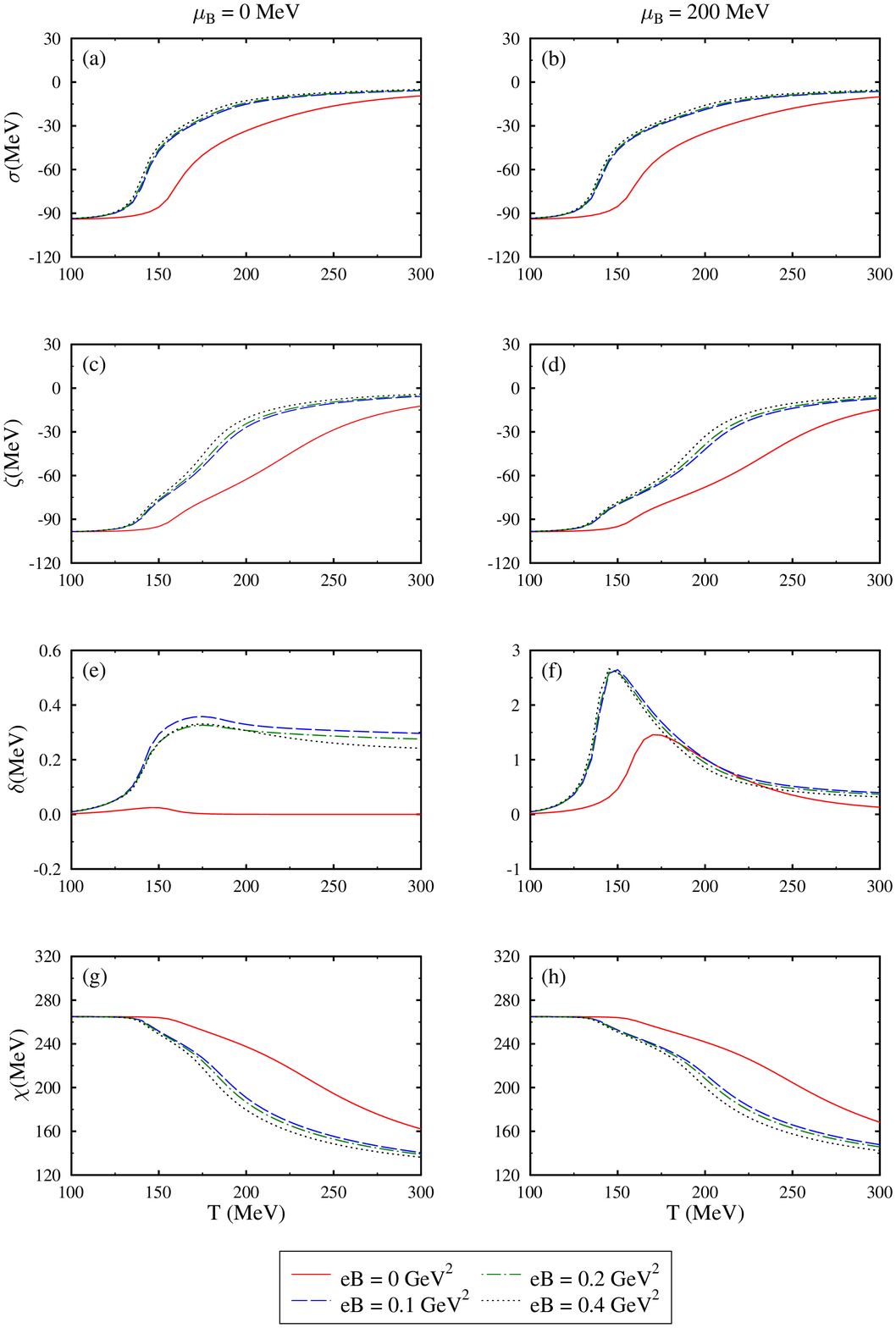}
    \caption{The scalar fields $\sigma$, $\zeta$, $\delta$ and $\chi$ are shown as a function of temperature T, for magnetic field, $eB$ = 0, 0.1, 0.2 and 0.4 GeV$^2$ and length of cubic volume, $R$ = $\infty$, baryonic chemical potential, $\mu_B$ = 0 and 200 MeV, isospin chemical potential, $\mu_I$ = 80 MeV and strangeness chemical potential, $\mu_S$ = 200 MeV.}
\label{figure1}    
\end{figure}

\begin{figure}
    \centering
    \includegraphics[scale=0.645]{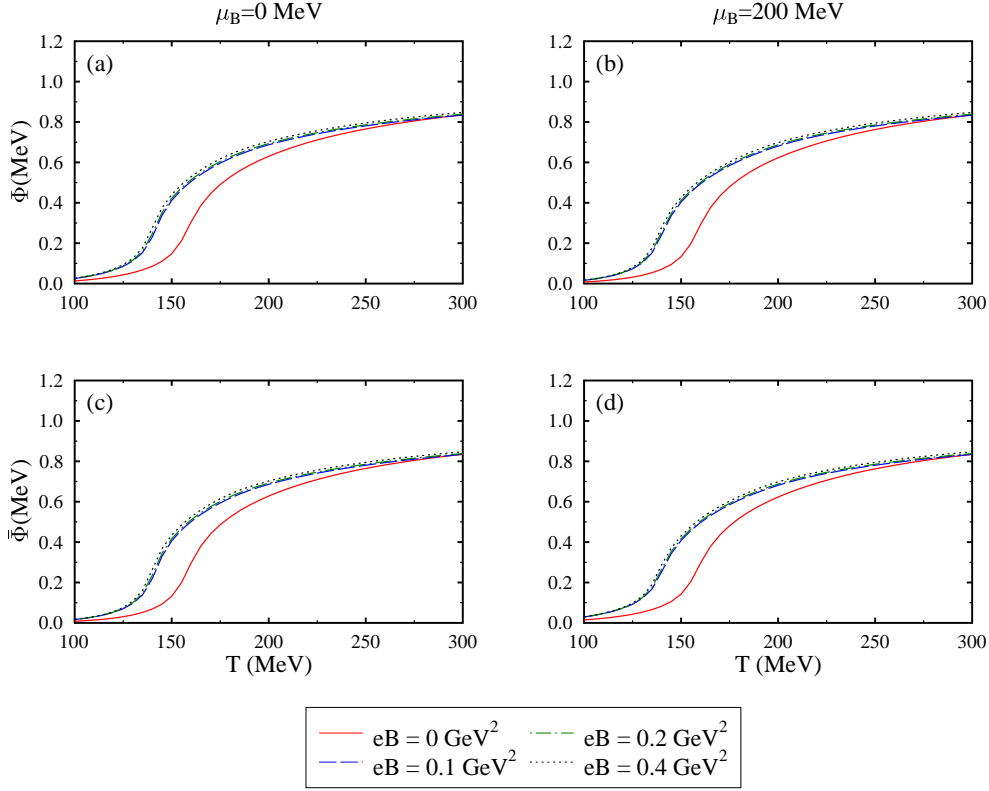}
    \caption{The Polyakov loop fields, $\Phi$ and $\bar{\Phi}$ shown as a function of temperature T, for magnetic field, $eB$ = 0, 0.1, 0.2 and 0.4 GeV$^2$ and length of cubic volume, $R$ = $\infty$, baryonic chemical potential, $\mu_B$ = 0 and 200 MeV, isospin chemical potential, $\mu_I$ = 80 MeV and strangeness chemical potential, $\mu_S$ = 200 MeV.}
    \label{figure2}
\end{figure}

\subsection{Thermodynamic properties and phase diagram}\label{thermosub}

\begin{figure}
    \centering
    \includegraphics[scale=0.610]{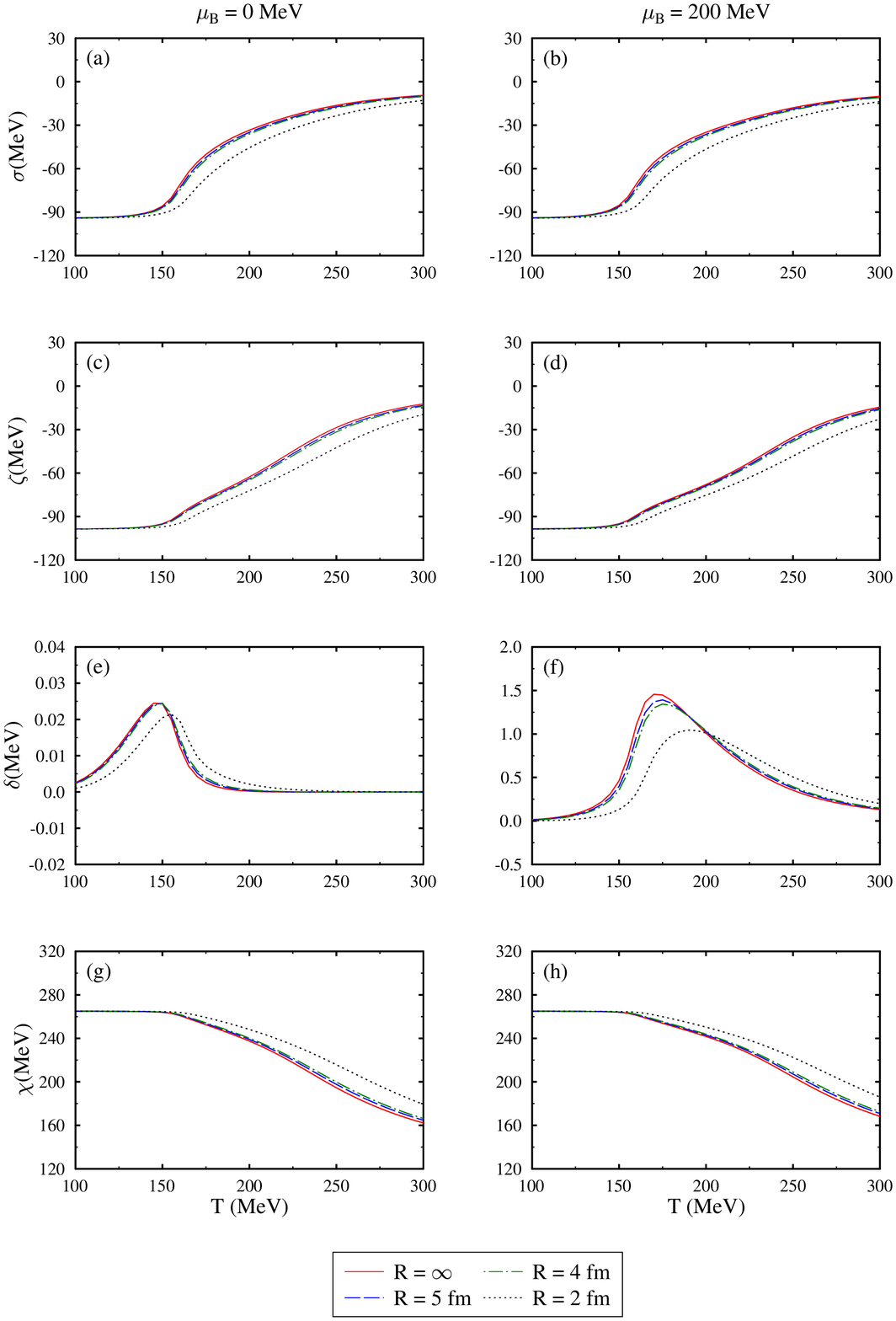}
    \caption{The scalar fields $\sigma$, $\zeta$, $\delta$ and $\chi$ shown as a function of temperature T, for length of cubic volume, $R$ = $\infty$, 5, 4 and 2 fm and magnetic field, $eB$ =0, baryonic chemical potential, $\mu_B$ = 0 and 200 MeV, isospin chemical potential, $\mu_I$ = 80 MeV and strangeness chemical potential, $\mu_S$ = 200 MeV.}
    \label{figure3}
\end{figure}

\begin{figure}
    \centering
    \includegraphics[scale=0.675]{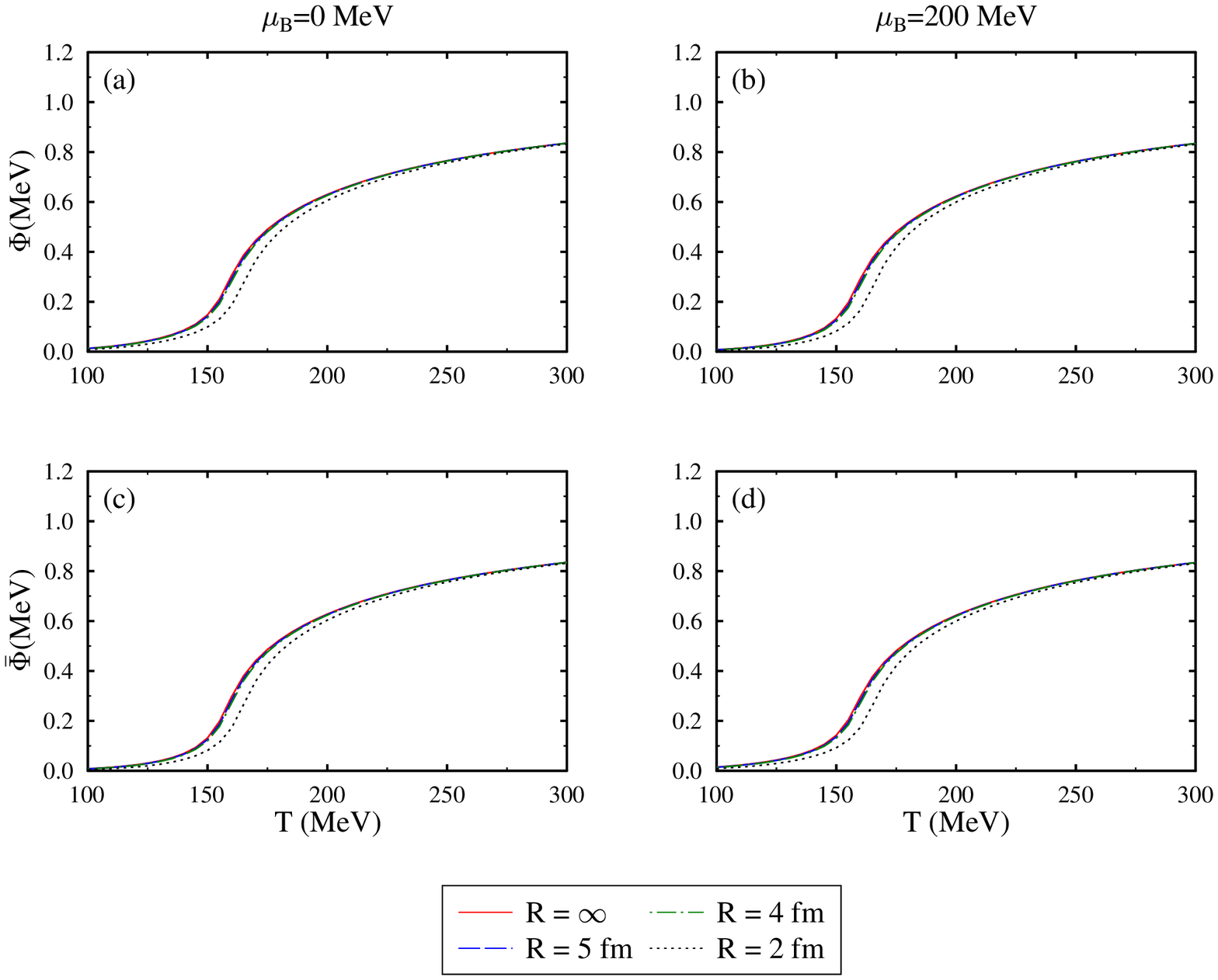}
    \caption{The Polyakov loop fields, $\Phi$ and $\bar{\Phi}$ shown as a function of temperature T, for length of cubic volume, $R$ = $\infty$, 5, 4 and 2 fm and magnetic field, $eB$ =0, baryonic chemical potential, $\mu_B$ = 0 and 200 MeV, isospin chemical potential, $\mu_I$ = 80 MeV and strangeness chemical potential, $\mu_S$ = 200 MeV.}
    \label{figure4}
\end{figure}

\begin{figure}
\centering
    \includegraphics[scale=0.675]{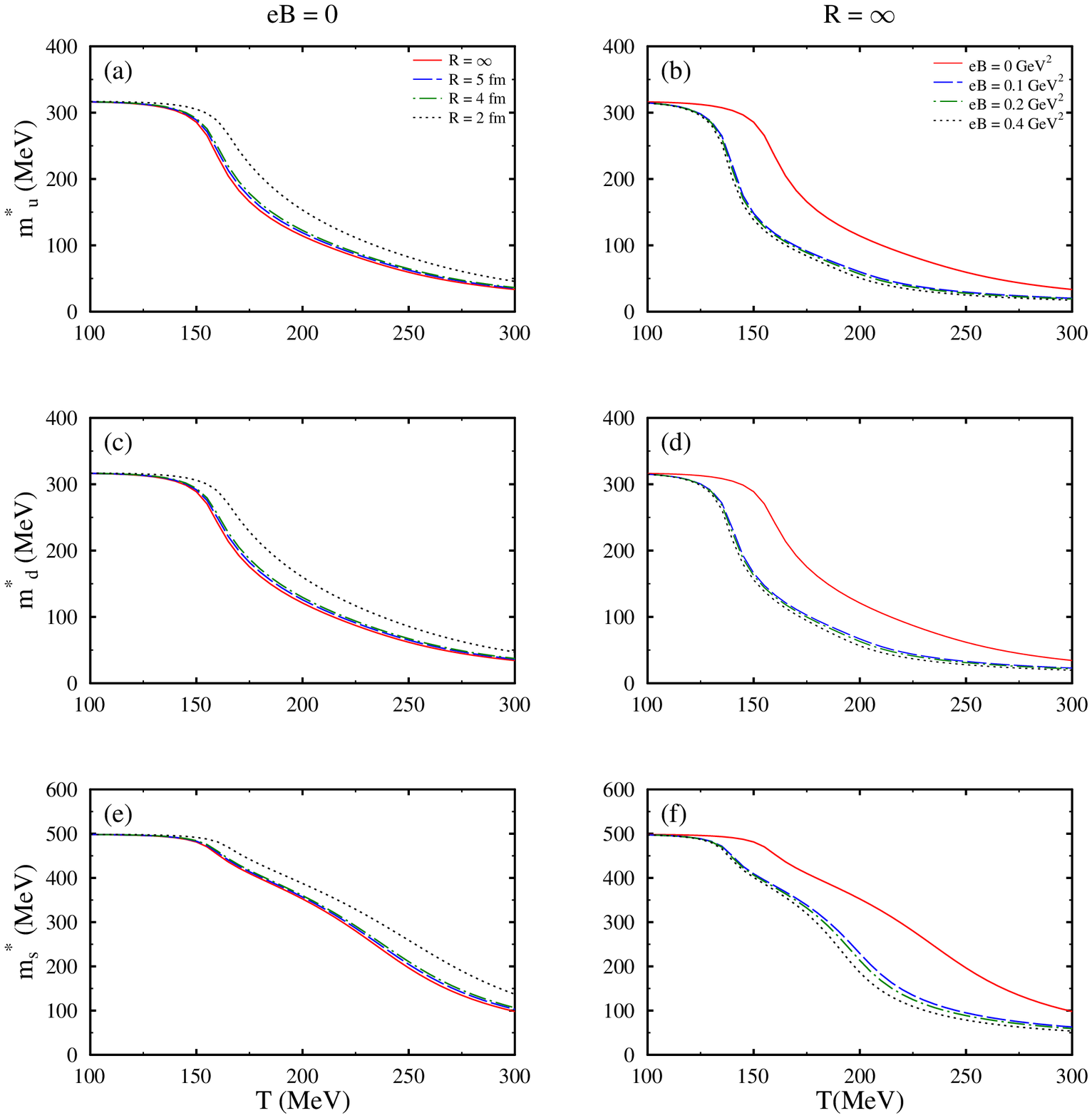}
    \caption{The effective quark masses $m_u^*$, $m_d^*$ and $m_s^*$ shown as a function of temperature T, for different values of length of cubic volume ($R$) and magnetic field ($eB$), baryonic chemical potential fixed at $\mu_B$ = 200 MeV, isospin chemical potential, $\mu_I$ = 80 MeV and strangeness chemical potential, $\mu_S$ = 200 MeV.}
    \label{figure5}
\end{figure}

\begin{figure}
    \centering
    \includegraphics[scale=0.675]{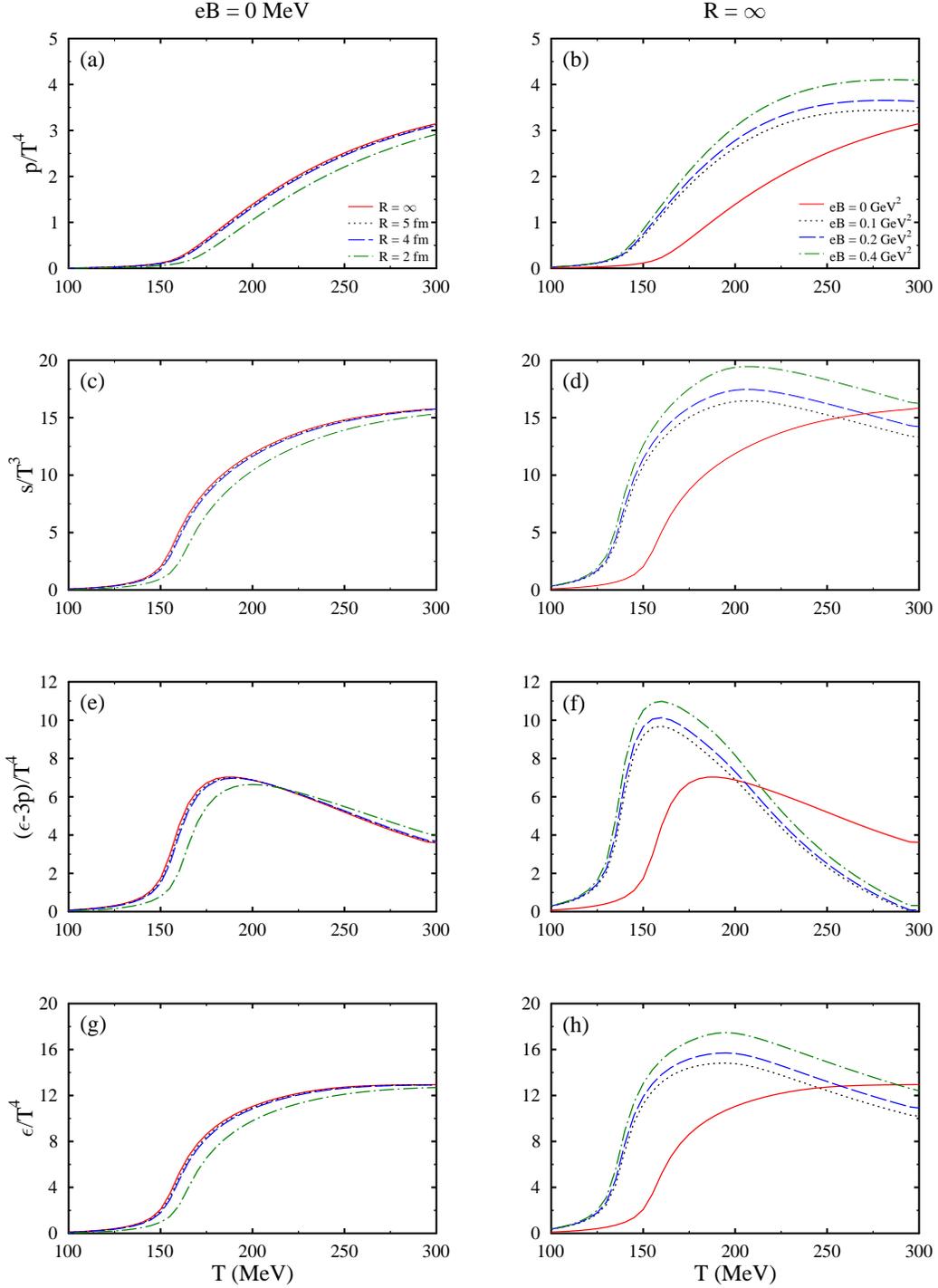}
    \caption{The pressure density, $p$, entropy density, $s$, trace anomaly ($\epsilon$-3$p$)/$T^4$ and energy density, $\epsilon$ as a function of temperature for different values of length of cubic volume ($R$) and magnetic field ($eB$) at baryonic chemical potential, $\mu_B$ = 200 MeV, isospin chemical potential, $\mu_I$ = 80 MeV and strangeness chemical potential, $\mu_S$ = 200 MeV. }
    \label{figure6}
\end{figure}

In this section, we have highlighted the impact of the pre-defined length of cubic volume, and external magnetic field on the chiral and deconfinement phase transitions of the Polyakov loop extended chiral quark mean field model. The asymmetry of the medium is assimilated by the induction of isospin chemical potential, $\mu_I$ = 80 MeV, and strangeness chemical potential, $\mu_S$, is fixed at 200 MeV from Figure~\ref{figure1} to Figure~\ref{figure6}.
In Figure~\ref{figure1} and Figure~\ref{figure2}, we have shown the variation of $\sigma$, $\zeta$, $\delta$, $\chi$ fields and Polyakov loop fields, $\Phi$ and $\bar{\Phi}$ as a function of temperature, T, for baryonic chemical potential, $\mu_B$ = 0 and 200 MeV and magnetic field, $eB$ = 0, 0.1, 0.2 and 0.4 GeV$^2$. We have noticed that for a given value of $\mu_B$, the value of $\sigma$ and $\zeta$ fields is constant till a specific temperature value is reached, after which the magnitude starts decreasing with the further increase in the temperature. The temperature at which we observe a sudden fall in the magnitude of scalar fields, is termed as pseudo-critical temperature, $T_p$ and is determined
from the inflection points of the scalar fields.
 For the scalar field $\sigma$ one inflection point is observed whereas for $\zeta$ field two points are found.
As we will see later, the position of inflection points will be determined from the derivatives of these scalar fields as a function of temperature.
 With the increase in the magnitude of the magnetic field for a given value of $\mu_B$ and T, we observe a drop in the magnitude of $\sigma$, $\zeta$ and $\chi$ field. As discussed in section~\ref{intro}, this slackening due to the external magnetic field is accredited to the suppression of quark condensate, referred as \enquote{inverse magnetic catalysis}.


At finite and zero values of $\mu_B$, it is clear that the value of $T_p$ is shifted to a lower temperature value. The magnitude of $\sigma$, $\zeta$, $\delta$ and $\chi $ fields increases with the increase in baryonic chemical potential for T $>$ 200 MeV. The scalar isovector field, $\delta$ induces the isospin asymmetry to the medium. In the quark matter, the magnitude of $\delta$ field is derived from the difference between the scalar densities of up and down quark. Hence, in Figure~\ref{figure1}(e), the value of $\delta$ is almost zero for the vanishing magnetic field and baryonic chemical potential. On the other side, we see non-zero magnitude of $\delta$ field for a finite value of baryonic chemical potential and vanishing magnetic field. The asymmetry in the medium at zero magnetic field value is introduced by finite isospin chemical potential. Due to the Landau quantization in the presence of the external magnetic field, we observe an enhancement in the magnitude of $\delta$ field for a given value of T and $\mu_B$. Figure~\ref{figure1}(g) and~\ref{figure1}(h) show the variation of the dilaton field, which incorporates the property of trace anomaly in the quark mean-field model. It shows the same behavior as $\sigma$ and $\zeta$ fields. It first remains constant up to a certain value of temperature and then decreases monotonically with a further temperature rise.

\begin{figure}
    \centering
    \includegraphics[scale=0.675]{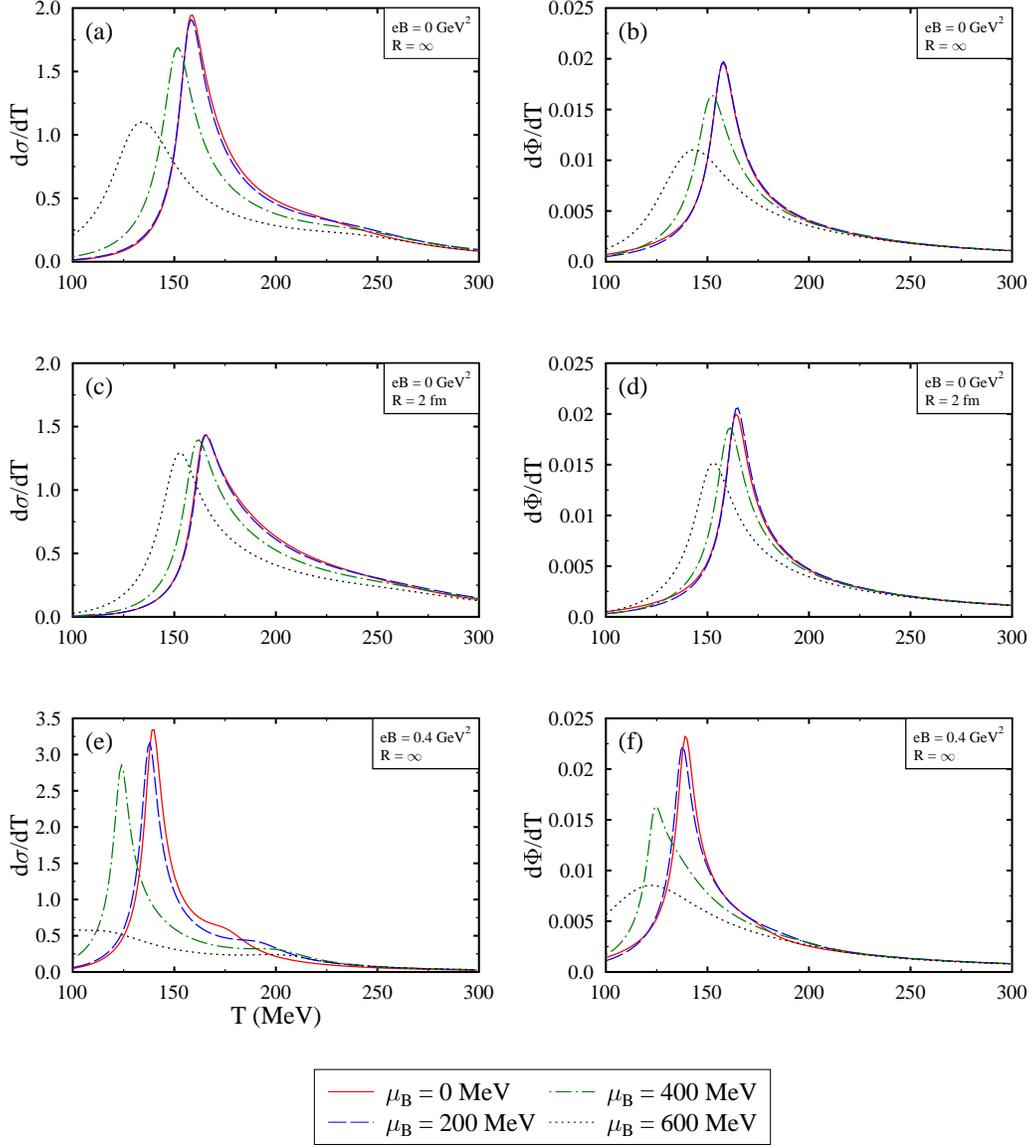}
    \caption{The derivative of scalar field $\sigma$ and Polyakov loop field, $\Phi$ shown as a function of temperature T, varying values of length of cubic volume, $R$ and magnetic field, $eB$. The baryonic chemical potential, $\mu_B$ = 0, 200, 400, 600 MeV, isospin chemical potential, $\mu_I$ = 80 MeV and strangeness chemical potential, $\mu_S$ = 200 MeV.}
    \label{figurenew}
\end{figure}

\begin{figure}
    \centering
    \includegraphics[scale=0.675]{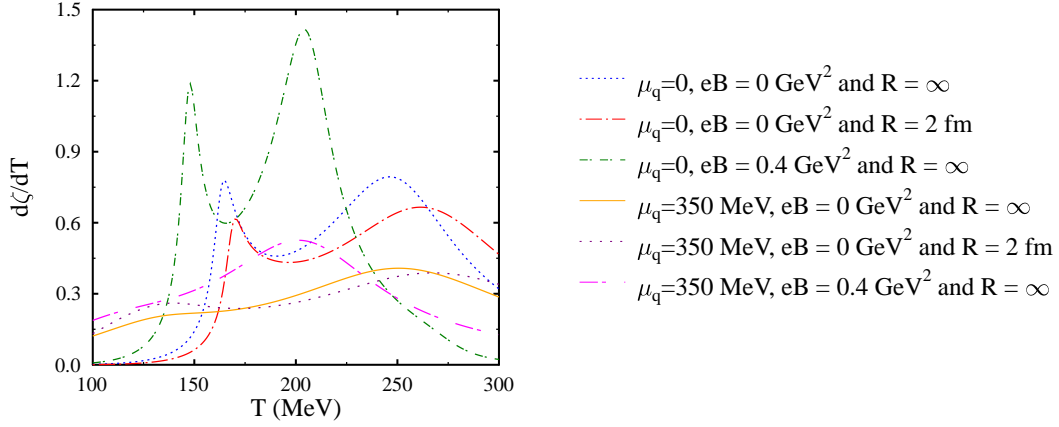}
    \caption{The derivative of strange scalar field $\zeta$ shown as a function of temperature T for $\mu_q$ = 0, 350 MeV at varying values of length of cubic volume $R$ and magnetic field, $eB$.}
    \label{figurezeta}
\end{figure}

\begin{figure}
    \centering
    \includegraphics[scale=0.675]{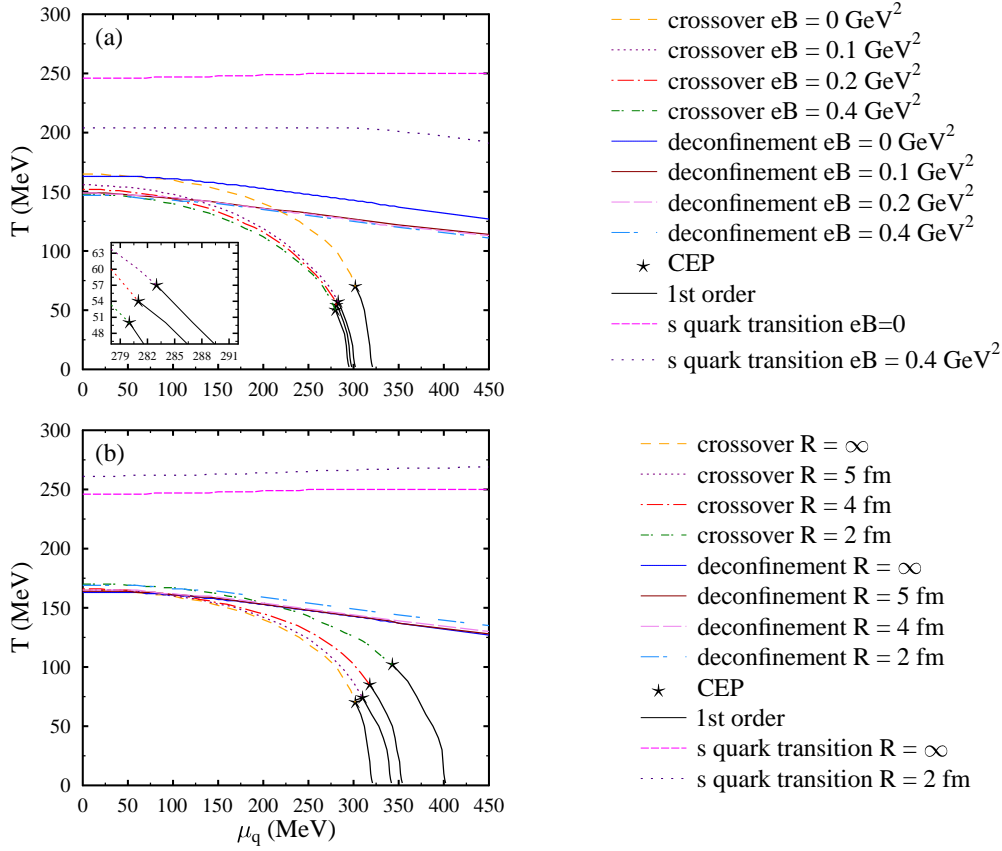}
    \caption{In the above panel, the $T-\mu_q$ phase diagram for $eB$ = 0, 0.1, 0.2 and 0.4 GeV$^2$ at infinite value of length of cubic volume ($R$) has been plotted. In the below panel, we show the phase diagram for various cubic length, $R$ = $\infty$, 5, 4 and 2 fm and zero value of magnetic field ($eB$).}
    \label{figure7}
\end{figure}

\begin{table}[b]
\centering
\begin{tabular}{|c|c|c|}
\hline
 $eB$ (GeV $^2$) &       $T_{CP}$ (MeV)  &     $\mu_{q(CP)}$ (MeV)     \\ \hline
   0                 & 70              & 302                           \\ \hline
0.1     &    57     &   283    \\ \hline
0.2      &     54    &   281    \\ \hline
0.4      &     50    &   280    \\ \hline
\end{tabular}
\caption{The value of critical-temperature and quark chemical potential for finite values of magnetic field.}
\label{table3}
\end{table}

\begin{table}[b]
\centering
\begin{tabular}{|c|c|c|}
\hline
 $R$ ($fm$) &       $T_{CP}$ (MeV)  &     $\mu_{q(CP)}$ (MeV)     \\ \hline
   $\infty$                & 70              & 302                           \\ \hline
 5    &    74     &   310    \\ \hline
4      &    85     &   318    \\ \hline
2      &     102    &   343    \\ \hline
\end{tabular}
\caption{The value of critical-temperature and quark chemical potential for finite values of system size.}
\label{table4}
\end{table}

In Figure~\ref{figure2}, for the vanishing value of baryonic chemical potential, both $\Phi$ and $\bar{\Phi}$ are same for varying temperature and magnetic field values. The value of $\Phi$ and $\bar{\Phi}$ is nearly zero in the low temperature region indicating the confined state for a given value of $\mu_B$. With an increase in the temperature, the value of $\Phi$ and $\bar{\Phi}$ increases due to the conversion from a confined hadronic state to a deconfined state. At non-zero magnetic field value, Polyakov loop fields show an increase for a given value of temperature and baryonic chemical potential. On further increasing the magnetic field, no significant change is observed. With the increase in the value of baryonic chemical potential, there is a slight decrease in values of deconfinement order parameters that might indicate the decrease in deconfinement transition temperature.

\begin{figure}
 
    \includegraphics[scale=0.675]{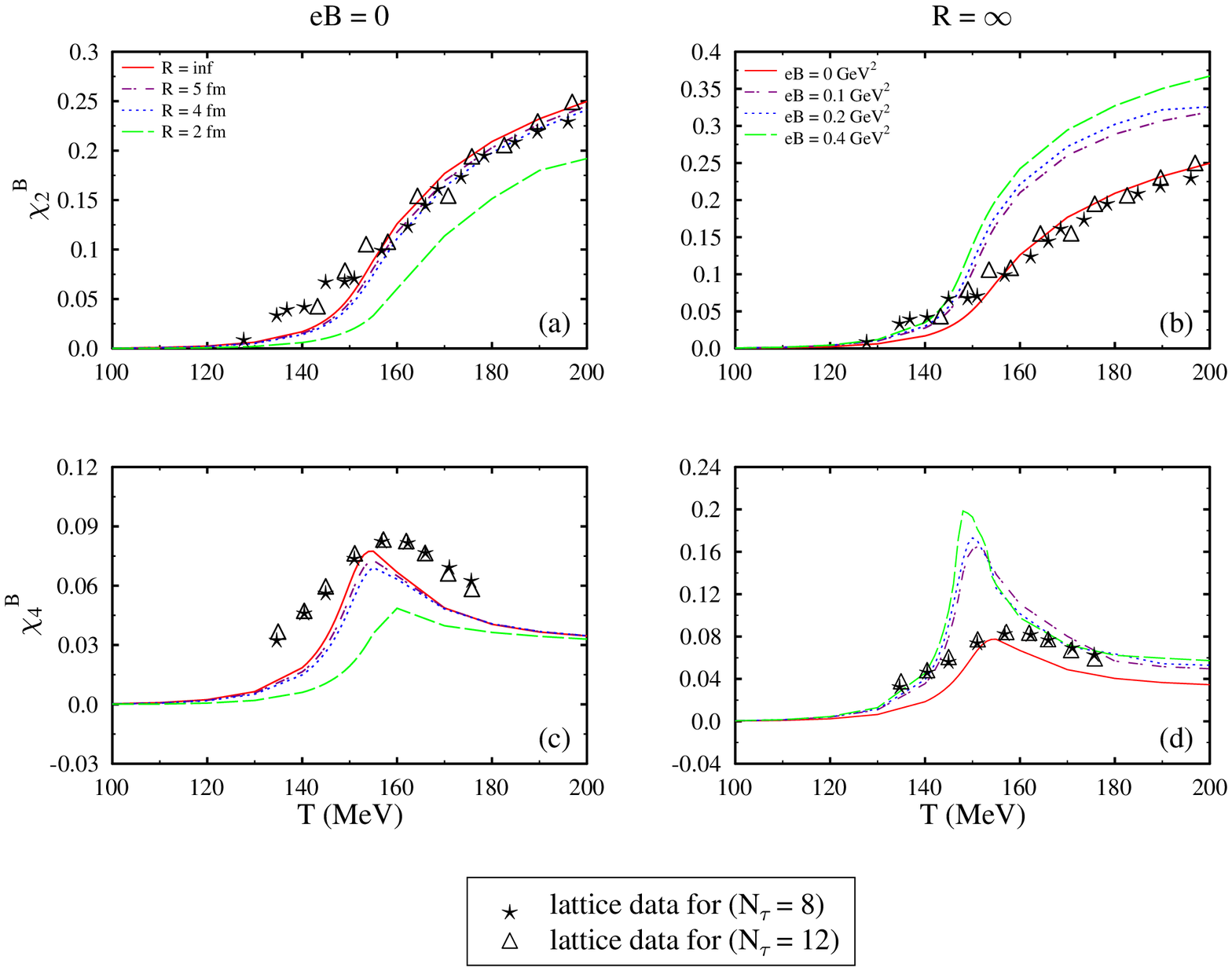}
    \caption{The second order baryon number susceptibility ($\chi_{2}^B$) and fourth order baryon number susceptibility ($\chi_{4}^B$) as a function of temperature for varying value of system size ($R$) and magnetic field ($eB$). The data has been compared with lattice data. }
    \label{figure8}
\end{figure}

\begin{figure}
    \centering
    \includegraphics[scale=0.675]{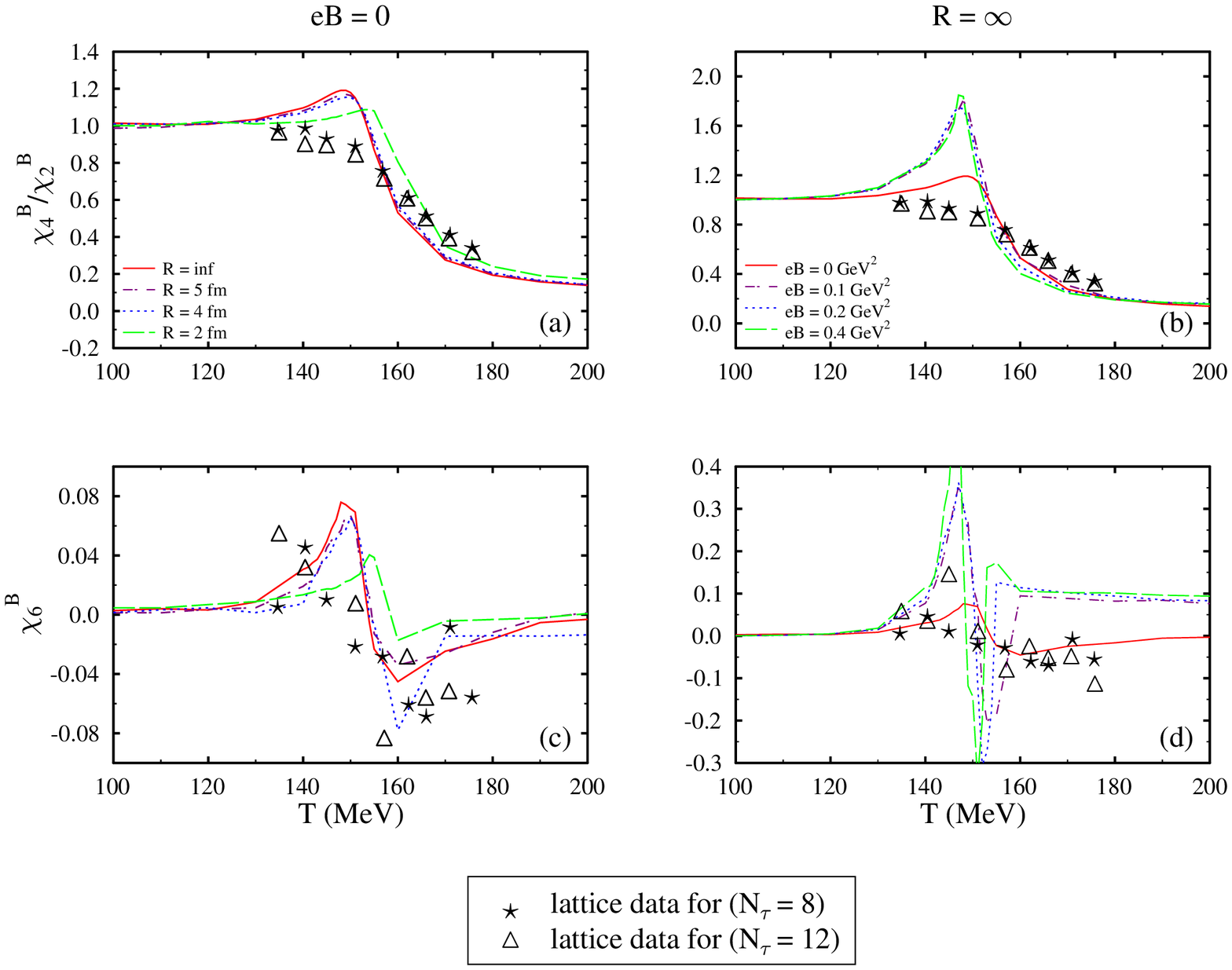}
    \caption{The kurtosis ($\chi_{4}^B$/$\chi_{2}^B$) and sixth order baryon number susceptibility,  ($\chi_{6}^B$) as a function of temperature for varying value of system size ($R$) and magnetic field ($eB$). The data has been compared with lattice data. }
    \label{figure9}
\end{figure}

Figure~\ref{figure3} and Figure~\ref{figure4} depict the variation of $\sigma$, $\zeta$, $\delta$, $\chi$ fields and Polyakov loop fields $\Phi$ and $\bar{\Phi}$ as a function of temperature, T, for baryonic chemical potential, $\mu_B$ = 0 and 200 MeV for varying system sizes. We observe that the magnitude of $\sigma$, $\zeta$ and $\chi$ fields increases with the decrease in the system volume for a given temperature value and baryonic chemical potential. The inflection point for these fields is shifted to higher temperature value with reducing system size, which may signify the shift of pseudo-critical temperature to higher values with a decrease in the system size. Similar results have been obtained for decreasing system volume by employing the PQM model in \cite{magdy2019influence,magdy2017influence}. The perceptible volume effects are observable only for $R \leq 5$ fm \cite{bernhardt2021critical}. There is no significant change in the $\delta $ field values for $R$ = 5, 4 fm because the scalar densities of $u$ and $d$ quarks remain almost same for zero value of baryonic chemical potential. Although, a shift of local maxima to higher temperature value for $R$ = 2 fm at finite value of isospin chemical potential is observed. The finite value of isospin potential here contributes the very low values of $\delta$ field and hence asymmetry. For finite value of $\mu_B$, we see an increase in the value of local maxima of $\delta$ field for a fixed system size. The maximum value of $\delta$ field decreases with the decrease in the system size for both finite and zero value of $\mu_B$ while the peak shifts to a higher value of temperature.

In Figure~\ref{figure4}, we perceive a fall in the magnitude of deconfinement parameters with decrease in system volume, which is in contrast to that observed with the finite magnetic field. An increase in the magnitude of $\Phi$ and $\bar{\Phi}$ is observed for increasing magnetic field. But this decrement in values of Polyakov loop conjugates for decreasing system size is more appreciable in case of $R$ = 2 fm.

\begin{figure}
    \centering
    \includegraphics[scale=0.675]{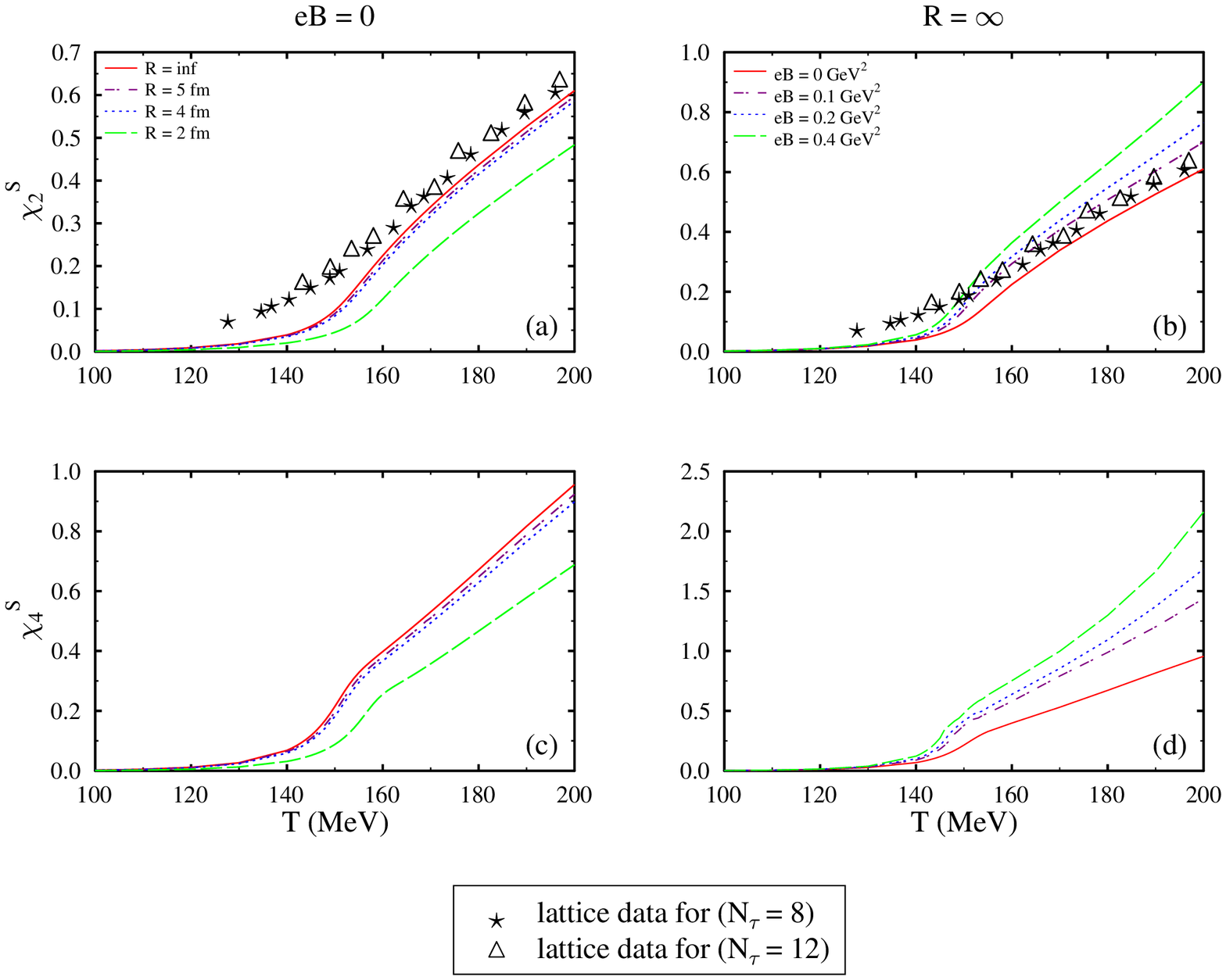}
    \caption{The second order strangeness number susceptibility ($\chi_{2}^S$) and fourth order strangeness number susceptibility ($\chi_{4}^S$) as a function of temperature for varying value of system size ($R$) and magnetic field ($eB$). The data has been compared with lattice data. }
    \label{figure10}
\end{figure}

Figure~\ref{figure5} depicts the modification of quark masses in the effective mean field as a function of temperature T for different values of length of cubic volume ($R$) and magnetic field ($eB$). The effective masses of quarks are derived from the coupling of scalar fields $\sigma$, $\zeta$ and $\delta$ with the quarks. As already discussed, the magnitude of scalar fields increases due to a decrease in the system volume, hence we observe an increase in the effective mass of quarks for a given value of temperature and chemical potential. On the other hand, due to the fall in magnitude of scalar fields with the increasing magnetic field, quark masses are reduced due to inverse magnetic catalysis. This happens as a result of the suppression of quark condensates with the increasing magnetic field. Also, the sudden fall in value of masses of quarks, signifying the change in degrees of freedom, seems to happen at a higher temperature with decreasing system size, whereas the pseudo-critical temperature shifts to a lower value of temperature with the rise in the magnetic field.

\begin{figure}
    \centering
    \includegraphics[scale=0.675]{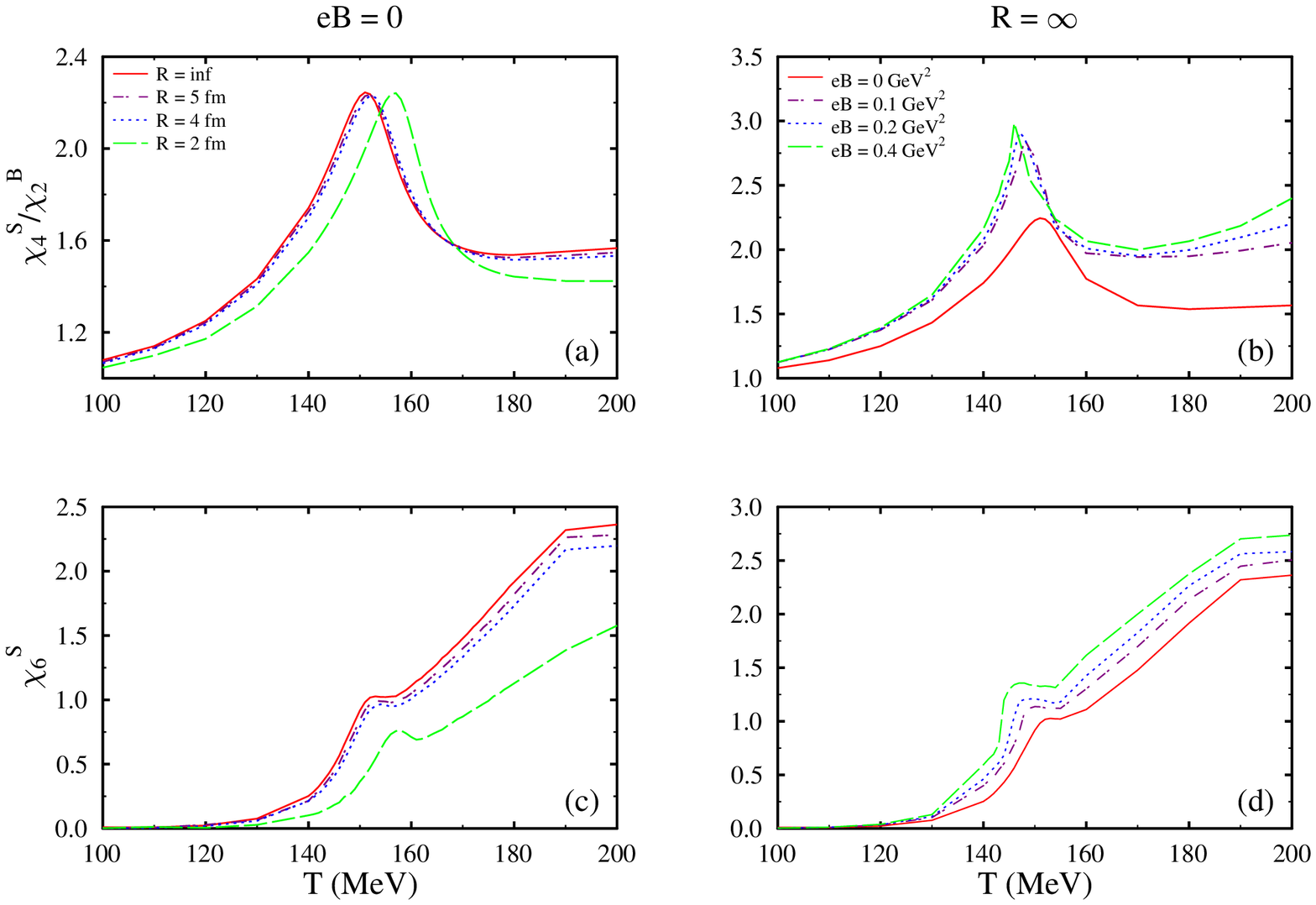}
    \caption{The kurtosis ($\chi_{4}^S$/$\chi_{2}^S$) and sixth order strangeness number susceptibility,  ($\chi_{6}^S$) as a function of temperature for varying value of system size ($R$) and magnetic field ($eB$).  }
    \label{figure11}
\end{figure}

 In Figure~\ref{figure6}, we have studied the variations of thermodynamic quantities, pressure density, $p$, entropy density, $s$, trace anomaly ($\epsilon$-3p)/$T^4$ and energy density, $\epsilon$ as a function of temperature for varying values of system size and magnetic field at the baryonic chemical potential $\mu_B$ fixed at 200 MeV.
 A sharp rise in the value of $p$, $s$, ($\epsilon$-3p)/$T^4$ and $\epsilon$ is noticed in the vicinity of the transition temperature. 
 A similar trend for thermodynamic quantities has been observed in many different studies \cite{tawfik20143,chahal2022quark,kumari2021quark}. The value of trace anomaly is quite small for low temperature values, due to the confinement of quarks. For high values of temperature, interaction strength becomes weaker and quarks are in the free state. The values increase sharply at transition temperature for all thermodynamic quantities and then advance to the ideal gas limit. The Stefann Boltzmann (SB) limit depends on the number of flavors considered in the study \cite{kumari2021quark,borsanyi2012fluctuations}. For decrease in the system size, there is a slight fall in pressure and entropy density values. A similar impact of finite volume on thermodynamic properties of quark matter has also been reported in \cite{bhattacharyya2013thermodynamic}. So we conclude that there is an enhancement in the thermodynamic quantities with the increase in magnetic field strength, while there is a reduction in values of thermodynamic properties for decreasing system volume for temperature range 100-240 MeV. There is a change in the trend for trace anomaly at higher values in case of varying system sizes. Also, the variation of entropy density, trace anomaly and energy density with the temperature at different magnetic field values changes for $T > 240 MeV$.

In Figure~\ref{figurenew}, we have plotted the derivative of non-strange scalar field $\sigma$ and Polyakov loop variable $\Phi$ as a function of temperature for varying magnetic fields and system sizes. There are three types of phase transitions in the QCD phase diagram, first is the chiral symmetry restoration of $u$ and $d$ quarks, second chiral symmetry restoration of $s$ quarks and the third is deconfinement transition. The value of critical temperatures for these three phase transition may be referred as $T_c^q$, $T_c^s$ and $T_c^d$, respectively. If the number of peaks is
one in temperature derivative of order parameters, then the critical temperature is obtained by the location of the peak. At zero value of the magnetic field and infinite system size, the value of $T_c^q  \approx  $ 160 MeV and $T_c^d \approx$ 158 MeV for vanishing values of baryonic chemical potential. When we further increase the baryonic chemical potential, the chiral transition temperature shifts to a lower temperature value. In all the cases, significant change in critical temperature value is observed for $\mu_B > 200$ MeV. For finite value of system size and zero magnetic field, we see a shift of peak to higher value of temperature for a given value of baryonic chemical potential. On the contrary, relocation of peak to lower temperature value is seen for finite value of magnetic field. A similar trend of chiral transition temperature and deconfinement transition temperature is perceived in all scenarios of magnetic field and system volume. 

In Figure~\ref{figurezeta}, we have plotted the derivative of strange scalar field $\zeta$ for different system sizes and magnetic field values. For the derivative of strange fields $\zeta$, two peaks are observed at zero value of quark chemical potential. The first peak
  coincides with the value of $T_c^q$ of light quarks whereas the second peak corresponds to the chiral phase transition for $s$ quark. When there are two peaks in temperature derivative of condensate, for the chiral phase transition, the critical temperature can be obtained by the
peak temperature analogous to $ \sigma(T) / \sigma(T = 0) < $ 1/2, while for deconfinement phase transition, it can be obtained by using the relation $\Phi$(T)/$\Phi(T \to \infty) >$ 1/2 \cite{mao2010phase}. At higher values of quark chemical potential, the first peak vanishes as in case of $\mu_q$ = 350 MeV.

In Figure~\ref{figure7}, we have plotted the QCD phase diagram at finite values of magnetic field and varying system sizes. 
The three types of phase transitions, i.e, the phase transition for light $u$ and $d$ quarks, the phase transition of $s$ quarks and deconfinement phase transition corresponding to Polyakov fields are shown in this figure.
For light quarks, at zero value of magnetic field and infinite system size (Figure~\ref{figure7}(a)), ($T_{CP}$,$\mu_{q(CP)}$) = (70,302) MeV. For infinite system size and $eB$ = 0.1 GeV$^2$, the value of $T_{CP}$ = 57 MeV and $\mu_{q(CP)}$ = 283 MeV. The critical point ($T_{CP}$,$\mu_{q(CP)}$) = (54,281) MeV for $eB$ = 0.2 GeV$^2$  and ($T_{CP}$,$\mu_{q(CP)}$) = (50,280) MeV for $eB$ = 0.4 GeV$^2$ for infinite system volume. The critical-point values for varying magnetic field strength are listed in Table~\ref{table3}. Hence we can conclude that critical point shifts to lower value of temperature and quark chemical potential due to increasing magnetic field. By Figure~\ref{figure7}(b), it is clear that with decrease in value of length of cubic volume, the critical-point moves towards higher value of temperature and quark chemical potential. For magnetic field, $eB$ = 0 and $R$ = 5 fm, the value of $T_{CP}$ = 74 MeV and $\mu_{q(CP)}$ = 310 MeV. The critical point ($T_{CP}$,$\mu_{q(CP)}$) = (85,318) MeV for $R$ = 4 fm and ($T_{CP}$,$\mu_{q(CP)}$) = (102,343) MeV for $R$ = 2 fm at zero value of the magnetic field. The critical-point values for different system sizes are listed in Table~\ref{table4}. This behavior of critical point is analogous to the one discussed in \cite{magdy2019influence,magdy2017influence} for PLSM. The deconfinement phase boundary is shifted to higher temperature values with decreasing system size and lower temperature value with increasing magnetic field strength. The chiral phase transition for $s$ quark and deconfinement transition is always a crossover in the QCD phase diagram \cite{costa2019phase}. On the other hand, for light quarks, the phase transition is a crossover till the critical-point and changes to first-order phase transition at higher values of quark chemical potential. The chiral phase boundary for $s$ quark follows the same trend as deconfinement phase transition in case of varying system size and magnetic field. The phase boundary is thus modified due to finite volume and finite magnetic field. It is to be noted that the value of isospin and strangeness chemical potential is considered to be zero while plotting the phase diagram. In earlier studies of the PCQMF model, the phase boundary was concluded to shift towards low values of temperature and higher values of quark chemical potential for increasing vector-interaction, whereas a relocation of critical point to lower temperature and quark chemical potential values have been reported for increasing isospin chemical potential \cite{chahal2022quark}.

\begin{figure}
    \centering
    \includegraphics[scale=0.675]{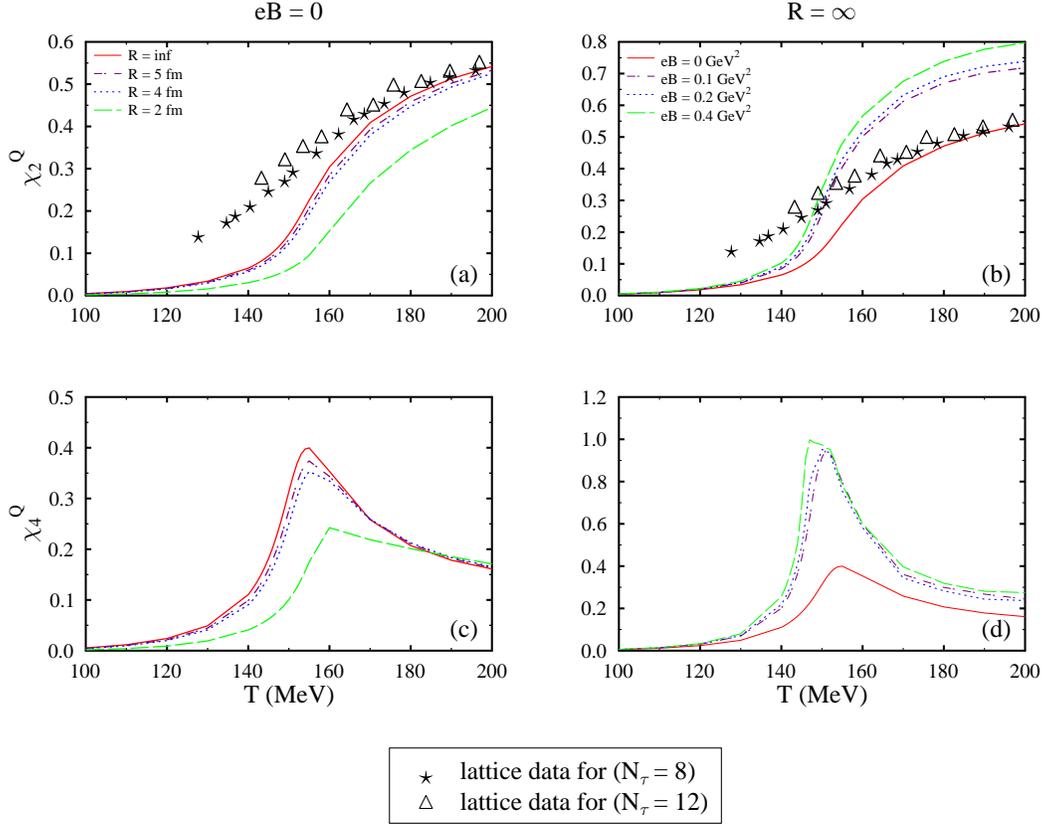}
    \caption{The second order charge number susceptibility ($\chi_{2}^Q$) and fourth order charge number susceptibility ($\chi_{4}^q$) as a function of temperature for varying values of system size ($R$) and magnetic field ($eB$). The data has been compared with lattice data. }
    \label{figure12}
\end{figure}

\subsection{Susceptibilities of conserved charges}\label{sussub}

In this section, we will discuss the susceptibilities of baryon number, strangeness number and charge number for varying values of magnetic field and finite system size. These susceptibilities have been calculated by expanding Taylor's series around $\mu_B = \mu_S = \mu_Q = 0$. Susceptibilities of conserved charges carry information about the QCD critical point and hence are important to study in the QCD phase diagram. It is due to the fact that above mentioned charges are conserved during the evolution of matter produced in heavy ion collisions and hence their fluctuations can be extracted by event-by-event analysis of the experiments \cite{abelev2009azimuthal,doi:10.1142/5029}. In Figure~\ref{figure8}, we have plotted the variation of the second and fourth-order baryon number susceptibility ($\chi_{2}^B$ and $\chi_{4}^B$) as a function of temperature for different values of system size ($R$) and magnetic field ($eB$). Susceptibilities have been calculated by using the Taylor series expansion discussed earlier in section~\ref{taylor} at vanishing chemical potential. The results obtained in the current study have been compared with lattice data for zero value of the magnetic field and infinite system size \cite{latticedata,2020lattice}. In the left panel, we can clearly see that value of $\chi_2^B$ and  $\chi_4^B$ decreases with the decrease in the system volume. As a function of T, a sudden increase in the value of susceptibilities is observed near the transition regime. This enhancement of susceptibilities may be an important signature of QCD critical point. This decrease in susceptibility of conserved charge due to decreasing system volume has also been observed in the PQM model \cite{magdy2019influence}. On the other hand, there is an increase in the value of susceptibilities with increase in value of magnetic field. This is attributed to the \enquote{inverse magnetic catalysis}, which have been discussed earlier. The susceptibilities of conserved charges in external magnetic field have been studied in \cite{fu2013fluctuations}. We observe that the thermodynamic quantities studied in the previous section show a rise with increment in external magnetic field but decreases with decreasing system size for $T < 220$ MeV. When the transition becomes an exact second order, the fluctuations of conserved charges must be divergent in nature \cite{fu2013fluctuations}.

In Figure~\ref{figure9}, we have plotted the kurtosis ($\chi_{4}^B$/$\chi_{2}^B$) and sixth order susceptibility ($\chi_{6}^B$) of baryon number. Kurtosis is considered as crucial observable in order to investigate the location of CEP owing to its sensitivity for both chiral and deconfinement transitions \cite{EJIRI2006275,STOKIC2009192}. In low temperature range, ($\chi_{4}^B$/$\chi_{2}^B$) is approximately one which represents the confined state of quarks whereas its value drops to $\approx$ 0.1 at high temperatures due to the change in degrees of freedom of quarks. It is not possible to determine skewness by using Taylor series expansion for zero value of baryonic chemical potential \cite{PhysRevD.73.114007}. This is due to the disintegration of odd terms in Taylor series expansion as a result of charge-parity symmetry. The ($\chi_{4}^B$/$\chi_{2}^B$) and ($\chi_{6}^B$) show a similar trend for finite values of magnetic field and volume. There is also a shift in the peak of kurtosis towards higher temperature as system size decreases which confirms the relocation of transition temperature towards higher values. On the other hand, we have opposite shift towards low temperatures with increasing magnetic field values.
\begin{figure}
    \centering
    \includegraphics[scale=0.675]{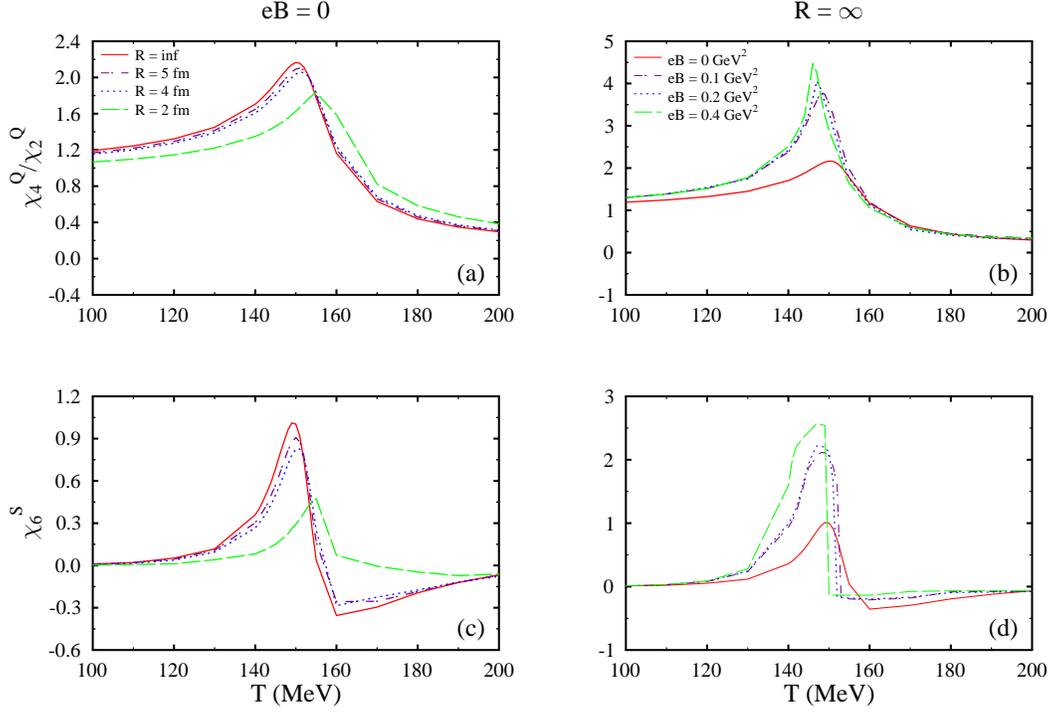}
    \caption{The kurtosis ($\chi_{4}^Q$/$\chi_{2}^Q$) and sixth order charge 
      number susceptibility, ($\chi_{6}^Q$) as a function of temperature for varying value of system size ($R$) and magnetic field ($eB$). }
    \label{figure13}
\end{figure}

Figure~\ref{figure10} and Figure~\ref{figure11}
display the variation of susceptibilities of strangeness number as a function of temperature. In the confined state with hadrons as degrees of freedom, mainly kaons contribute to the strangeness of system. This happens because of the suppressed production of heavy strange mesons in this phase \cite{PhysRevLett.95.182301}. But in the deconfined phase, strangeness is contributed by low mass hadrons. This results in the maximization of fluctuations of strangeness number and hence plays a vital role in investigating deconfinement transition \cite{borsanyi2012fluctuations}. The trend of $\chi_{2}^S$ is similar to that of $\chi_{2}^B$, it increases monotonically with increase in temperature. It has been observed in PNJL model that impact of varying magnetic field is more pronounced in case of $s$ quarks, especially for low order fluctuations \cite{fu2013fluctuations}. A peaked structure is observed in the kurtosis of strangeness number in the transition regime. We observe a sharp rise in the second, fourth and sixth order strangeness number susceptibilities in temperature range of 140-160 MeV, signalling the change of phase from confined hadronic to deconfined QGP phase.
 In Figure~\ref{figure12} and Figure~\ref{figure13} we have shown charge number fluctuations as a temperature function. The second order susceptibilities of all conserved charges follow the similar smooth trend with increasing values of temperature. There is a considerable rise in value of susceptibility value near the phase transition region with increasing magnetic field whereas the magnitude decreases with decreasing system size. Also there is a shift towards lower temperature for different magnetic field values but an opposite dislocation to higher temperature is observed by reducing system volume. The effect of external magnetic field on the fluctuations of conserved charges have been shown to enhance the magnitude of susceptibilities in the transition regime in the PNJL model \cite{fu2013fluctuations}. All the above results are in  fair agreement with the lattice QCD simulations for $\chi_{2}^Q$ at high temperature. 
 
\section{Summary} \label{summary}
We have analyzed the impact of finite volume and external magnetic field on the thermodynamic properties using Polyakov loop extended chiral SU(3) quark mean field model in the asymmetric quark matter. The impact of external magnetic field and finite system size on the phase diagram of QCD have been investigated by inspecting the variation of scalar and vector fields along with Polyakov loop variables at different values of temperature and chemical potential. Various thermodynamic properties like pressure density, energy density, entropy density, trace anomaly property and quark masses have been studied for varying system size and magnetic field. We have observed an increase in magnitude of  thermodynamic quantities studied in the current work with the increasing magnetic field in temperature range 125-250 MeV whereas there is a fall in the magnitude with decreasing system size. Phase boundary is found to shift towards lower values of temperature and quark chemical potential with increasing magnetic field and higher value of temperature and quark chemical potential for decreasing system volume. The fluctuations of conserved charges, baryon number, strangeness number and charge number have been calculated for various values of magnetic field and length of cubic volume. Susceptibilities of conserved charges are found to be enhanced in the regime of critical-point. The peaked structure of quartic susceptibilities becomes more and more pronounced with the rise in magnetic field strength. These fluctuations can be deduced from event-by-event inspection of the experimental data and hence play a significant role in the determination of CEP. The results obtained have been compared with the lattice QCD simulations data for zero value of magnetic field and infinite system size.
 In future work, we will calculate the fluctuations of conserved charges for varying values of magnetic field and system volume for non-zero value of chemical potential. The functional renormalization approach can be employed to study thermodynamic variables of quark matter beyond mean-field \cite{schaefer2008renormalization}. The work will be extended in the future to study the higher order cumulants using automatic differentiation \cite{mathias}.

 \section*{Acknowledgment}

Authors sincerely acknowledge the support towards this work from the Ministry of Science and Human Resources (MHRD), Government of India via Institute fellowship under the National Institute of Technology Jalandhar. Arvind Kumar sincerely acknowledges the DST-SERB, Government of India for funding of research project CRG/2019/000096.

\bibliographystyle{elsarticle-num}

\bibliography{bib2}

\begin{thebibliography}{100}
\expandafter\ifx\csname url\endcsname\relax
  \def\url#1{\texttt{#1}}\fi
\expandafter\ifx\csname urlprefix\endcsname\relax\def\urlprefix{URL }\fi
\expandafter\ifx\csname href\endcsname\relax
  \def\href#1#2{#2} \def\path#1{#1}\fi

\bibitem{busza2018heavy}
W.~Busza, K.~Rajagopal, W.~Van Der~Schee, Heavy ion collisions: the big
  picture, and the big questions, arXiv preprint arXiv:1802.04801 (2018).

\bibitem{ohnishi2012phase}
A.~Ohnishi, Phase diagram and heavy-ion collisions: overview, Progress of
  Theoretical Physics Supplement 193 (2012) 1--10.

\bibitem{pandav2022search}
A.~Pandav, D.~Mallick, B.~Mohanty, Search for the qcd critical point in high
  energy nuclear collisions, Progress in Particle and Nuclear Physics (2022)
  103960.

\bibitem{kumar2013star}
{L. Kumar and STAR Collaboration}, Star results from the rhic beam energy
  scan-i, Nuclear Physics A 904 (2013) 256c--263c.

\bibitem{odyniec2010rhic}
G.~Odyniec, Rhic beam energy scan program—experimental approach to the qcd
  phase diagram, Journal of Physics G: Nuclear and Particle Physics 37~(9)
  (2010) 094028.

\bibitem{bruning2012large}
O.~Br{\"u}ning, H.~Burkhardt, S.~Myers, The large hadron collider, Progress in
  Particle and Nuclear Physics 67~(3) (2012) 705--734.

\bibitem{sissakian2009nuclotron}
{A. Sissakian, A. Sorin and NICA collaboration}, The nuclotron-based ion
  collider facility (nica) at jinr: new prospects for heavy ion collisions and
  spin physics, Journal of Physics G: Nuclear and Particle Physics 36~(6)
  (2009) 064069.

\bibitem{durante2019all}
M.~Durante, P.~Indelicato, B.~Jonson, V.~Koch, K.~Langanke, U.-G. Mei{\ss}ner,
  E.~Nappi, T.~Nilsson, T.~St{\"o}hlker, E.~Widmann, et~al., All the fun of the
  fair: fundamental physics at the facility for antiproton and ion research,
  Physica Scripta 94~(3) (2019) 033001.

\bibitem{stephanov2006qcd}
M.~Stephanov, Qcd phase diagram: An overview, arXiv preprint hep-lat/0701002
  (2006).

\bibitem{detar2009qcd}
C.~DeTar, U.~Heller, Qcd thermodynamics from the lattice, The European Physical
  Journal A 41~(3) (2009) 405--437.
\newblock \href {https://doi.org/https://doi.org/10.1140/epja/i2009-10825-3}
  {\path{doi:https://doi.org/10.1140/epja/i2009-10825-3}}.

\bibitem{gupta1999lattice}
R.~Gupta, Lattice qcd, in: AIP Conference Proceedings CONF-981188, Vol. 490,
  American Institute of Physics, 1999, pp. 3--9.

\bibitem{aoki2006order}
Y.~Aoki, G.~Endr{\H{o}}di, Z.~Fodor, S.~D. Katz, K.~K. Szab{\'o}, The order of
  the quantum chromodynamics transition predicted by the standard model of
  particle physics, Nature 443~(7112) (2006) 675--678.

\bibitem{guenther2021overview}
J.~N. Guenther, Overview of the qcd phase diagram, The European Physical
  Journal A 57~(4) (2021) 1--23.

\bibitem{goy2017sign}
V.~Goy, V.~Bornyakov, D.~Boyda, A.~Molochkov, A.~Nakamura, A.~Nikolaev,
  V.~Zakharov, Sign problem in finite density lattice qcd, Progress of
  Theoretical and Experimental Physics 2017~(3) (2017) 031D01.

\bibitem{muroya2003lattice}
S.~Muroya, A.~Nakamura, C.~Nonaka, T.~Takaishi, Lattice qcd at finite density:
  an introductory review, Progress of theoretical physics 110~(4) (2003)
  615--668.

\bibitem{danzer2009study}
J.~Danzer, C.~Gattringer, L.~Liptak, M.~Marinkovic, A study of the sign problem
  for lattice qcd with chemical potential, Physics Letters B 682~(2) (2009)
  240--245.

\bibitem{fodor2002new}
Z.~Fodor, S.~D. Katz, A new method to study lattice qcd at finite temperature
  and chemical potential, Physics Letters B 534~(1-4) (2002) 87--92.

\bibitem{ejiri2008canonical}
S.~Ejiri, Canonical partition function and finite density phase transition in
  lattice qcd, Physical Review D 78~(7) (2008) 074507.

\bibitem{hatta2003universality}
Y.~Hatta, T.~Ikeda, Universality, the qcd critical and tricritical point, and
  the quark number susceptibility, Physical Review D 67~(1) (2003) 014028.

\bibitem{fodor2004critical}
Z.~Fodor, S.~D. Katz, Critical point of qcd at finite t and $\mu$, lattice
  results for physical quark masses, Journal of High Energy Physics 2004~(04)
  (2004) 050.

\bibitem{stephanov2004qcd}
M.~Stephanov, Qcd phase diagram and the critical point, Progress of Theoretical
  Physics Supplement 153 (2004) 139--156.

\bibitem{lacey2015indications}
R.~A. Lacey, Indications for a critical end point in the phase diagram for hot
  and dense nuclear matter, Physical Review Letters 114~(14) (2015) 142301.

\bibitem{tlusty2018rhic}
D.~Tlusty, The rhic beam energy scan phase ii: physics and upgrades, arXiv
  preprint arXiv:1810.04767 (2018).

\bibitem{stephanov2005qcd}
M.~Stephanov, Qcd critical point and correlations, in: Journal of Physics:
  Conference Series, Vol.~27, IOP Publishing, 2005, p. 144.

\bibitem{garg2013conserved}
P.~Garg, D.~Mishra, P.~Netrakanti, B.~Mohanty, A.~Mohanty, B.~Singh, N.~Xu,
  Conserved number fluctuations in a hadron resonance gas model, Physics
  Letters B 726~(4-5) (2013) 691--696.

\bibitem{maris2003dyson}
P.~Maris, C.~D. Roberts, Dyson--schwinger equations: a tool for hadron physics,
  International Journal of Modern Physics E 12~(03) (2003) 297--365.

\bibitem{papazoglou1998chiral}
P.~Papazoglou, S.~Schramm, J.~Schaffner-Bielich, H.~Stoecker, W.~Greiner,
  Chiral lagrangian for strange hadronic matter, Physical Review C 57~(5)
  (1998) 2576.

\bibitem{bowler1995nonlocal}
R.~D. Bowler, M.~Birse, A nonlocal, covariant generalisation of the njl model,
  Nuclear Physics A 582~(3-4) (1995) 655--664.

\bibitem{schaefer2007phase}
B.-J. Schaefer, J.~M. Pawlowski, J.~Wambach, Phase structure of the
  polyakov-quark-meson model, Physical Review D 76~(7) (2007) 074023.

\bibitem{kashiwa2008critical}
K.~Kashiwa, H.~Kouno, M.~Matsuzaki, M.~Yahiro, Critical endpoint in the
  polyakov-loop extended njl model, Physics Letters B 662~(1) (2008) 26--32.

\bibitem{dupuis2021nonperturbative}
N.~Dupuis, L.~Canet, A.~Eichhorn, W.~Metzner, J.~M. Pawlowski, M.~Tissier,
  N.~Wschebor, The nonperturbative functional renormalization group and its
  applications, Physics Reports 910 (2021) 1--114.

\bibitem{harrison1973origin}
E.~Harrison, Origin of magnetic fields in the early universe, Physical Review
  Letters 30~(5) (1973) 188.

\bibitem{enqvist1993primordial}
K.~Enqvist, P.~Olesen, On primordial magnetic fields of electroweak origin,
  Physics Letters B 319~(1-3) (1993) 178--185.

\bibitem{inghirami2020magnetic}
G.~Inghirami, M.~Mace, Y.~Hirono, L.~Del~Zanna, D.~E. Kharzeev, M.~Bleicher,
  Magnetic fields in heavy ion collisions: flow and charge transport, The
  European Physical Journal C 80~(3) (2020) 1--26.

\bibitem{kharzeev2008effects}
D.~E. Kharzeev, L.~D. McLerran, H.~J. Warringa, The effects of topological
  charge change in heavy ion collisions:“event by event p and cp
  violation”, Nuclear Physics A 803~(3-4) (2008) 227--253.

\bibitem{deng2012event}
W.-T. Deng, X.-G. Huang, Event-by-event generation of electromagnetic fields in
  heavy-ion collisions, Physical Review C 85~(4) (2012) 044907.

\bibitem{vachaspati1991magnetic}
T.~Vachaspati, Magnetic fields from cosmological phase transitions, Physics
  Letters B 265~(3-4) (1991) 258--261.

\bibitem{vallgren2011amorphous}
C.~Y. Vallgren, G.~Arduini, J.~Bauche, S.~Calatroni, P.~Chiggiato, K.~Cornelis,
  P.~C. Pinto, B.~Henrist, E.~M{\'e}tral, H.~Neupert, et~al., Amorphous carbon
  coatings for the mitigation of electron cloud in the cern super proton
  synchrotron, Physical Review Special Topics-Accelerators and Beams 14~(7)
  (2011) 071001.

\bibitem{fu2013fluctuations}
W.-j. Fu, Fluctuations and correlations of hot qcd matter in an external
  magnetic field, Physical Review D 88~(1) (2013) 014009.

\bibitem{skokov2009estimate}
V.~Skokov, A.~Y. Illarionov, V.~Toneev, Estimate of the magnetic field strength
  in heavy-ion collisions, International Journal of Modern Physics A 24~(31)
  (2009) 5925--5932.

\bibitem{abelev2009azimuthal}
B.~Abelev, M.~Aggarwal, Z.~Ahammed, A.~Alakhverdyants, B.~Anderson,
  D.~Arkhipkin, G.~Averichev, J.~Balewski, O.~Barannikova, L.~Barnby, et~al.,
  Azimuthal charged-particle correlations and possible local strong parity
  violation, Physical review letters 103~(25) (2009) 251601.

\bibitem{fukushima2010chiral}
K.~Fukushima, M.~Ruggieri, R.~Gatto, Chiral magnetic effect in the
  polyakov--nambu--jona-lasinio model, Physical Review D 81~(11) (2010) 114031.

\bibitem{balinew2012qcd}
G.~S. Bali, F.~Bruckmann, G.~Endr{\H{o}}di, Z.~Fodor, S.~Katz, A.~Sch{\"a}fer,
  Qcd quark condensate in external magnetic fields, Physical Review D 86~(7)
  (2012) 071502.

\bibitem{bruckmann2013inverse}
F.~Bruckmann, G.~Endr{\H{o}}di, T.~G. Kovacs, Inverse magnetic catalysis and
  the polyakov loop, Journal of High Energy Physics 2013~(4) (2013) 1--23.

\bibitem{shushpanov1997quark}
I.~Shushpanov, A.~V. Smilga, Quark condensate in a magnetic field, Physics
  Letters B 402~(3-4) (1997) 351--358.

\bibitem{klimenko1992three}
K.~Klimenko, Three-dimensional gross-neveu model at nonzero temperature and in
  the presence of an external electromagnetic field, Zeitschrift f{\"u}r Physik
  C Particles and Fields 54~(2) (1992) 323--329.

\bibitem{shovkovy2013magnetic}
I.~A. Shovkovy, Magnetic catalysis: a review, Strongly Interacting Matter in
  Magnetic Fields (2013) 13--49.

\bibitem{bali2013thermodynamic}
G.~Bali, F.~Bruckmann, M.~Constantinou, M.~Costa, G.~Endrodi, Z.~Fodor,
  S.~Katz, S.~Krieg, H.~Panagopoulos, A.~Schafer, K.~Szabo, Thermodynamic
  properties of qcd in external magnetic fields, Proceedings of Science 171
  (2013) 0197.

\bibitem{klevansky1989chiral}
S.~Klevansky, R.~H. Lemmer, Chiral-symmetry restoration in the
  nambu--jona-lasinio model with a constant electromagnetic field, Physical
  Review D 39~(11) (1989) 3478.

\bibitem{endrHodi2013qcd}
G.~Endr{\H{o}}di, Qcd equation of state at nonzero magnetic fields in the
  hadron resonance gas model, Journal of High Energy Physics 2013~(4) (2013)
  1--19.

\bibitem{bali2012qcd}
G.~Bali, F.~Bruckmann, G.~Endr{\H{o}}di, Z.~Fodor, S.~Katz, S.~Krieg,
  A.~Sch{\"a}fer, K.~Szabo, The qcd phase diagram for external magnetic fields,
  Journal of high energy physics 2012~(2) (2012) 1--25.

\bibitem{fraga2011finite}
E.~S. Fraga, L.~F. Palhares, P.~Sorensen, Finite-size scaling as a tool in the
  search for the qcd critical point in heavy ion data, Physical Review C 84~(1)
  (2011) 011903.

\bibitem{bzdak2020mapping}
A.~Bzdak, S.~Esumi, V.~Koch, J.~Liao, M.~Stephanov, N.~Xu, Mapping the phases
  of quantum chromodynamics with beam energy scan, Physics Reports 853 (2020)
  1--87.

\bibitem{luo2017search}
X.~Luo, N.~Xu, Search for the qcd critical point with fluctuations of conserved
  quantities in relativistic heavy-ion collisions at rhic: an overview, Nuclear
  Science and Techniques 28~(8) (2017) 1--40.

\bibitem{bhattacharyya2017polyakov}
A.~Bhattacharyya, S.~K. Ghosh, R.~Ray, K.~Saha, S.~Upadhaya,
  Polyakov--nambu--jona-lasinio model in finite volumes, EPL (Europhysics
  Letters) 116~(5) (2017) 52001.

\bibitem{fisher1972scaling}
M.~E. Fisher, M.~N. Barber, Scaling theory for finite-size effects in the
  critical region, Physical Review Letters 28~(23) (1972) 1516.

\bibitem{brezin1985finite}
E.~Br{\'e}zin, J.~Zinn-Justin, Finite size effects in phase transitions,
  Nuclear Physics B 257 (1985) 867--893.

\bibitem{magdy2019influence}
N.~Magdy, Influence of finite volume effect on the polyakov quark--meson model,
  Universe 5~(4) (2019) 94.

\bibitem{magdy2017influence}
N.~Magdy, M.~Csan{\'a}d, R.~A. Lacey, Influence of finite volume and magnetic
  field effects on the qcd phase diagram, Journal of Physics G: Nuclear and
  Particle Physics 44~(2) (2017) 025101.

\bibitem{mata2022effects}
N.~B. Mata~Carrizal, E.~Valbuena~Ord{\'o}{\~n}ez, A.~J. Garza~Aguirre, F.~J.
  Betancourt~Sotomayor, J.~R. Morones~Ibarra, Effects of a finite volume in the
  phase structure of qcd, Universe 8~(5) (2022) 264.

\bibitem{grunfeld2018finite}
A.~G. Grunfeld, G.~Lugones, Finite size effects in strongly interacting matter
  at zero chemical potential from polyakov loop nambu-jona-lasinio model in the
  light of lattice data, The European Physical Journal C 78~(8) (2018) 1--13.

\bibitem{kiriyama2005color}
O.~Kiriyama, Color-superconducting strangelets in the nambu--jona-lasinio
  model, Physical Review D 72~(5) (2005) 054009.

\bibitem{zhao2019chiral}
Y.-P. Zhao, R.-R. Zhang, H.~Zhang, H.-S. Zong, Chiral phase transition from the
  dyson-schwinger equations in a finite spherical volume, Chinese Physics C
  43~(6) (2019) 063101.

\bibitem{zhao2020finite}
Y.-P. Zhao, P.-L. Yin, Z.-H. Yu, H.-S. Zong, Finite volume effects on chiral
  phase transition and pseudoscalar mesons properties from the
  polyakov-nambu-jona-lasinio model, Nuclear Physics B 952 (2020) 114919.

\bibitem{lugones2019surface}
G.~Lugones, A.~G. Grunfeld, Surface tension of hot and dense quark matter under
  strong magnetic fields, Physical Review C 99~(3) (2019) 035804.

\bibitem{borsanyi2012fluctuations}
S.~Borsanyi, Z.~Fodor, S.~D. Katz, S.~Krieg, C.~Ratti, K.~Szabo, Fluctuations
  of conserved charges at finite temperature from lattice qcd, Journal of High
  Energy Physics 2012~(1) (2012) 1--15.

\bibitem{adam2020net}
J.~Adam, Net-proton number fluctuations and the quantum chromodynamics critical
  point, Tech. rep., Brookhaven National Lab.(BNL), Upton, NY (United States)
  (2020).

\bibitem{chahal2022quark}
N.~Chahal, S.~Dutt, A.~Kumar, Quark matter properties and fluctuations of
  conserved charges in (2+ 1)-flavored quark model, Chinese Physics C (2022).

\bibitem{chatterjee2012fluctuations}
S.~Chatterjee, K.~A. Mohan, Fluctuations and correlations of conserved charges
  in the (2+ 1) polyakov quark meson model, Physical Review D 86~(11) (2012)
  114021.

\bibitem{fan2019probing}
W.~Fan, X.~Luo, H.~Zong, Probing the qcd phase structure with higher order
  baryon number susceptibilities within the njl model, Chinese Physics C 43~(3)
  (2019) 033103.

\bibitem{asakawa2016fluctuations}
M.~Asakawa, M.~Kitazawa, Fluctuations of conserved charges in relativistic
  heavy ion collisions: An introduction, Progress in Particle and Nuclear
  Physics 90 (2016) 299--342.

\bibitem{berg20133}
B.~A. Berg, H.~Wu, Su (3) deconfining phase transition with finite volume
  corrections due to a confined exterior, Physical Review D 88~(7) (2013)
  074507.

\bibitem{cristoforetti2010thermodynamics}
M.~Cristoforetti, T.~Hell, B.~Klein, W.~Weise, Thermodynamics and quark
  susceptibilities: A monte carlo approach to the polyakov--nambu--jona-lasinio
  model, Physical Review D 81~(11) (2010) 114017.

\bibitem{ding2021fluctuations}
H.-T. Ding, S.-T. Li, Q.~Shi, X.-D. Wang, Fluctuations and correlations of net
  baryon number, electric charge and strangeness in a background magnetic
  field, The European Physical Journal A 57~(6) (2021) 1--13.

\bibitem{wang2003strange}
P.~Wang, V.~E. Lyubovitskij, T.~Gutsche, A.~Faessler, Strange quark matter in a
  chiral su (3) quark mean field model, Physical Review C 67~(1) (2003) 015210.

\bibitem{poberezhnyuk2017quantum}
R.~Poberezhnyuk, V.~Vovchenko, D.~Anchishkin, M.~Gorenstein, Quantum van der
  waals and walecka models of nuclear matter, International Journal of Modern
  Physics E 26~(10) (2017) 1750061.

\bibitem{wang2002multi}
P.~Wang, H.~Guo, Z.~Zhang, Y.~Yu, R.~Su, H.~Song, Multi-strange hadronic
  systems in a chiral su (3) quark mean field model, Nuclear Physics A
  705~(3-4) (2002) 455--474.

\bibitem{wang2004new}
P.~Wang, D.~Leinweber, A.~Thomas, A.~Williams, New treatment of the chiral su
  (3) quark mean field model, Nuclear Physics A 744 (2004) 273--292.

\bibitem{wang2001strange}
P.~Wang, Z.~Zhang, Y.~Yu, R.~Su, Q.~Song, Strange hadronic matter in a chiral
  su (3) quark mean-field model, Nuclear Physics A 688~(3-4) (2001) 791--807.

\bibitem{fukushima2017polyakov}
K.~Fukushima, V.~Skokov, Polyakov loop modeling for hot qcd, Progress in
  Particle and Nuclear Physics 96 (2017) 154--199.

\bibitem{hansen2020quark}
H.~Hansen, R.~Stiele, P.~Costa, Quark and polyakov-loop correlations in
  effective models at zero and nonvanishing density, Physical Review D 101~(9)
  (2020) 094001.

\bibitem{ratti2006phases}
C.~Ratti, M.~A. Thaler, W.~Weise, Phases of qcd: Lattice thermodynamics and a
  field theoretical model, Physical Review D 73~(1) (2006) 014019.

\bibitem{herbst2011phase}
T.~K. Herbst, J.~M. Pawlowski, B.-J. Schaefer, The phase structure of the
  polyakov--quark--meson model beyond mean field, Physics Letters B 696~(1-2)
  (2011) 58--67.

\bibitem{fu2021high}
W.-j. Fu, X.~Luo, J.~M. Pawlowski, F.~Rennecke, R.~Wen, S.~Yin, High-order
  baryon number fluctuations within the frg approach, Physical Review D 094047
  (2021) 104.

\bibitem{bando1988nonlinear}
M.~Bando, T.~Kugo, K.~Yamawaki, Nonlinear realization and hidden local
  symmetries, Physics Reports 164~(4-5) (1988) 217--314.

\bibitem{weinberg1968nonlinear}
S.~Weinberg, Nonlinear realizations of chiral symmetry, Physical Review 166~(5)
  (1968) 1568.

\bibitem{kharzeev2009broken}
D.~Kharzeev, E.~Levin, K.~Tuchin, Broken scale invariance, massless dilaton and
  confinement in qcd, Journal of High Energy Physics 2009~(06) (2009) 055.

\bibitem{beekman2019introduction}
A.~Beekman, L.~Rademaker, J.~van Wezel, An introduction to spontaneous symmetry
  breaking, SciPost Physics Lecture Notes (2019) 011.

\bibitem{kumari2021quark}
M.~Kumari, A.~Kumar, Quark matter within polyakov chiral su(3) quark mean field
  model at finite temperature, The European Physical Journal Plus 136~(1)
  (2021) 19.

\bibitem{papazoglou1999nuclei}
P.~Papazoglou, D.~Zschiesche, S.~Schramm, J.~Schaffner-Bielich, H.~St{\"o}cker,
  W.~Greiner, Nuclei in a chiral su(3) model, Physical Review C 59~(1) (1999)
  411.

\bibitem{PhysRevD.81.074013}
B.-J. Schaefer, M.~Wagner, J.~Wambach, Thermodynamics of ($2+1$)-flavor qcd:
  Confronting models with lattice studies, Phys. Rev. D 81 (2010) 074013.
\newblock \href {https://doi.org/10.1103/PhysRevD.81.074013}
  {\path{doi:10.1103/PhysRevD.81.074013}}.

\bibitem{PhysRevD.75.034007}
S.~Rossner, C.~Ratti, W.~Weise, Polyakov loop, diquarks, and the two-flavor
  phase diagram, Phys. Rev. D 75 (2007) 034007.
\newblock \href {https://doi.org/10.1103/PhysRevD.75.034007}
  {\path{doi:10.1103/PhysRevD.75.034007}}.

\bibitem{fukugita1990finite}
M.~Fukugita, M.~Okawa, A.~Ukawa, Finite-size scaling study of the deconfining
  phase transition in pure su (3) lattice gauge theory, Nuclear Physics B
  337~(1) (1990) 181--232.

\bibitem{tawfik20143}
A.~N. Tawfik, N.~Magdy, Su (3) polyakov linear-$\sigma$ model in an external
  magnetic field, Physical Review C 90~(1) (2014) 015204.

\bibitem{ferreira2014inverse}
M.~Ferreira, P.~Costa, O.~Louren{\c{c}}o, T.~Frederico, C.~Provid{\^e}ncia,
  Inverse magnetic catalysis in the (2+ 1)-flavor nambu--jona-lasinio and
  polyakov--nambu--jona-lasinio models, Physical Review D 89~(11) (2014)
  116011.

\bibitem{tawfik20183}
A.~N. Tawfik, A.~M. Diab, M.~Hussein, Su (3) polyakov linear-sigma model:
  magnetic properties of qcd matter in thermal and dense medium, Journal of
  Experimental and Theoretical Physics 126~(5) (2018) 620--632.

\bibitem{bhattacharyya2013thermodynamic}
A.~Bhattacharyya, P.~Deb, S.~K. Ghosh, R.~Ray, S.~Sur, Thermodynamic properties
  of strongly interacting matter in a finite volume using the
  polyakov--nambu--jona-lasinio model, Physical Review D 87~(5) (2013) 054009.

\bibitem{bhattacharyya2015thermodynamics}
A.~Bhattacharyya, R.~Ray, S.~Samanta, S.~Sur, Thermodynamics and fluctuations
  of conserved charges in a hadron resonance gas model in a finite volume,
  Physical Review C 91~(4) (2015) 041901.

\bibitem{skokov2010vacuum}
V.~Skokov, B.~Friman, E.~Nakano, K.~Redlich, B.-J. Schaefer, Vacuum
  fluctuations and the thermodynamics of chiral models, Physical Review D
  82~(3) (2010) 034029.

\bibitem{mao2010phase}
H.~Mao, J.~Jin, M.~Huang, Phase diagram and thermodynamics of the polyakov
  linear sigma model with three quark flavors, Journal of Physics G: Nuclear
  and Particle Physics 37~(3) (2010) 035001.

\bibitem{chatterjee2012including}
S.~Chatterjee, K.~A. Mohan, Including the fermion vacuum fluctuations in the
  (2+ 1) flavor polyakov quark-meson model, Physical Review D 85~(7) (2012)
  074018.

\bibitem{gupta2012revisiting}
U.~S. Gupta, V.~K. Tiwari, Revisiting the phase structure of the
  polyakov-quark-meson model in the presence of vacuum fermion fluctuation,
  Physical Review D 85~(1) (2012) 014010.

\bibitem{ping2001nuclear}
W.~Ping, Z.~Zong-Ye, Y.~You-Wen, Nuclear matter in a chiral su (3) quark
  mean-field model, Communications in Theoretical Physics 36~(1) (2001) 71.

\bibitem{shao2018baryon}
G.-y. Shao, Z.-d. Tang, X.-y. Gao, W.-b. He, Baryon number fluctuations and the
  phase structure in the pnjl model, The European Physical Journal C 78~(2)
  (2018) 1--7.
\newblock \href
  {https://doi.org/https://doi.org/10.1140/epjc/s10052-018-5636-0}
  {\path{doi:https://doi.org/10.1140/epjc/s10052-018-5636-0}}.

\bibitem{cheng2009baryon}
M.~Cheng, P.~Hegde, C.~Jung, F.~Karsch, O.~Kaczmarek, E.~Laermann,
  R.~Mawhinney, C.~Miao, P.~Petreczky, C.~Schmidt, et~al., Baryon number,
  strangeness, and electric charge fluctuations in qcd at high temperature,
  Physical Review D 79~(7) (2009) 074505.

\bibitem{skokov2011charge}
V.~Skokov, B.~Friman, F.~Karsch, K.~Redlich, Charge fluctuations in chiral
  models and the qcd phase transition, Journal of Physics G: Nuclear and
  Particle Physics 38~(12) (2011) 124102.

\bibitem{isserstedt2020dyson}
P.~Isserstedt, M.~Buballa, C.~S. Fischer, P.~J. Gunkel, Dyson-schwinger
  approach to baryon number fluctuations, in: Journal of Physics: Conference
  Series, Vol. 1667, IOP Publishing, 2020, p. 012015.

\bibitem{xin2014quark}
X.-y. Xin, S.-x. Qin, Y.-x. Liu, et~al., Quark number fluctuations at finite
  temperature and finite chemical potential via the dyson-schwinger equation
  approach, Physical Review D 90~(7) (2014) 076006.

\bibitem{PhysRevD.73.114007}
S.~K. Ghosh, T.~K. Mukherjee, M.~G. Mustafa, R.~Ray, Susceptibilities and speed
  of sound from the polyakov-nambu-jona-lasinio model, Phys. Rev. D 73 (2006)
  114007.
\newblock \href {https://doi.org/10.1103/PhysRevD.73.114007}
  {\path{doi:10.1103/PhysRevD.73.114007}}.

\bibitem{schmidt2010net}
C.~Schmidt, Net-baryon number fluctuations in (2+ 1)-flavor qcd, Progress of
  Theoretical Physics Supplement 186 (2010) 563--566.
\newblock \href {https://doi.org/https://doi.org/10.1143/PTPS.186.563}
  {\path{doi:https://doi.org/10.1143/PTPS.186.563}}.

\bibitem{bernhardt2021critical}
J.~Bernhardt, C.~S. Fischer, P.~Isserstedt, B.-J. Schaefer, Critical endpoint
  of qcd in a finite volume, Physical Review D 104~(7) (2021) 074035.

\bibitem{costa2019phase}
P.~Costa, R.~Pereira, Phase diagram, scalar-pseudoscalar meson behavior and
  restoration of symmetries in (2+ 1) polyakov-nambu-jona-lasinio model,
  Symmetry 11~(4) (2019) 507.

\bibitem{doi:10.1142/5029}
R.~C. Hwa, X.-N. Wang,
  \href{https://www.worldscientific.com/doi/abs/10.1142/5029}{Quark–Gluon
  Plasma 3}, WORLD SCIENTIFIC, 2004.
\newblock \href
  {http://arxiv.org/abs/https://www.worldscientific.com/doi/pdf/10.1142/5029}
  {\path{arXiv:https://www.worldscientific.com/doi/pdf/10.1142/5029}}, \href
  {https://doi.org/10.1142/5029} {\path{doi:10.1142/5029}}.
\newline\urlprefix\url{https://www.worldscientific.com/doi/abs/10.1142/5029}

\bibitem{latticedata}
A.~Bazavov, T.~Bhattacharya, M.~Cheng, C.~DeTar, H.-T. Ding, S.~Gottlieb,
  R.~Gupta, P.~Hegde, U.~M. Heller, F.~Karsch, E.~Laermann, L.~Levkova,
  S.~Mukherjee, P.~Petreczky, C.~Schmidt, R.~A. Soltz, W.~Soeldner, R.~Sugar,
  D.~Toussaint, W.~Unger, P.~Vranas, Chiral and deconfinement aspects of the
  qcd transition, Phys. Rev. D 85 (2012) 054503.
\newblock \href {https://doi.org/10.1103/PhysRevD.85.054503}
  {\path{doi:10.1103/PhysRevD.85.054503}}.

\bibitem{2020lattice}
A.~Bazavov, D.~Bollweg, H.-T. Ding, P.~Enns, J.~Goswami, P.~Hegde,
  O.~Kaczmarek, F.~Karsch, R.~Larsen, S.~Mukherjee, et~al., Skewness, kurtosis,
  and the fifth and sixth order cumulants of net baryon-number distributions
  from lattice qcd confront high-statistics star data, Physical Review D
  101~(7) (Apr 2020).
\newblock \href {https://doi.org/10.1103/physrevd.101.074502}
  {\path{doi:10.1103/physrevd.101.074502}}.

\bibitem{EJIRI2006275}
S.~Ejiri, F.~Karsch, K.~Redlich, Hadronic fluctuations at the qcd phase
  transition, Physics Letters B 633~(2) (2006) 275--282.
\newblock \href
  {https://doi.org/https://doi.org/10.1016/j.physletb.2005.11.083}
  {\path{doi:https://doi.org/10.1016/j.physletb.2005.11.083}}.

\bibitem{STOKIC2009192}
B.~Stokić, B.~Friman, K.~Redlich, Kurtosis and compressibility near the chiral
  phase transition, Physics Letters B 673~(3) (2009) 192--196.
\newblock \href
  {https://doi.org/https://doi.org/10.1016/j.physletb.2009.02.018}
  {\path{doi:https://doi.org/10.1016/j.physletb.2009.02.018}}.

\bibitem{PhysRevLett.95.182301}
V.~Koch, A.~Majumder, J.~Randrup, Baryon-strangeness correlations: A diagnostic
  of strongly interacting matter, Phys. Rev. Lett. 95 (2005) 182301.
\newblock \href {https://doi.org/10.1103/PhysRevLett.95.182301}
  {\path{doi:10.1103/PhysRevLett.95.182301}}.

\bibitem{schaefer2008renormalization}
B.-J. Schaefer, J.~Wambach, Renormalization group approach towards the qcd
  phase diagram, Physics of Particles and Nuclei 39~(7) (2008) 1025--1032.

\bibitem{mathias}
M.~Wagner, A.~Walther, B.-J. Schaefer, On the efficient computation of
  high-order derivatives for implicitly defined functions, Computer Physics
  Communications 181~(4) (2010) 756–764.
\newblock \href {https://doi.org/10.1016/j.cpc.2009.12.008}
  {\path{doi:10.1016/j.cpc.2009.12.008}}.

\end{thebibliography}

\end{document}